\documentclass[sigconf]{acmart}
\setcopyright{none}
\renewcommand\footnotetextcopyrightpermission[1]{}
\usepackage{natbib}
\usepackage{svg}
\usepackage{multirow}
\usepackage[colorinlistoftodos]{todonotes}
\usepackage[table,dvipsnames]{xcolor}

\colorlet{mylinkcolor}{YellowOrange}
\colorlet{mycitecolor}{Aquamarine}
\colorlet{myurlcolor}{violet}

\usepackage{fancyhdr}

\AtBeginDocument{%
  }

\newif\ifshowcomments

\usepackage{cleveref}
\usepackage{booktabs}
\usepackage{caption}
\usepackage{placeins}
\usepackage{bm}
\usepackage{bbm}
\usepackage[nolist]{acronym} 
\newacro{ASR}{Attack Success Rate}

\newcounter{logicrule}

\newcommand{\logicrule}[1]{%
  \refstepcounter{logicrule}%
  \par\smallskip
  \noindent
  \begin{minipage}{\linewidth}
    \small\textbf{Logic Rule~\thelogicrule.} #1
  \end{minipage}
  \par\medskip
}

\theoremstyle{definition}

\AtBeginDocument{%
    \hypersetup{ 
        colorlinks=true, 
        linkcolor=mylinkcolor, 
        citecolor=mycitecolor, 
        urlcolor=myurlcolor, 
        filecolor=myurlcolor }
}

\usepackage[inline]{enumitem}
\newenvironment{inlinelist}{\begin{enumerate*}[label=\emph{(\roman{*})}]}{\end{enumerate*}}

\usepackage{makecell}

\begin{document}

\title[Does Reasoning Mitigate Backdoor Attacks? A Neuro-Symbolic Perspective]{Does Reasoning Mitigate Backdoor Attacks?\\A Neuro-Symbolic Perspective}

\thispagestyle{fancy}
\fancyhf{}
\fancyhead[C]{Manuscript under review}
\renewcommand{\headrulewidth}{0pt}

\author{Marco Antonio Corallo}
\orcid{0009-0007-4854-9421}
\correspondingauthor
\affiliation{%
  \institution{\textit{{\"O}rebro University}}
  \city{{\"O}rebro}
  \country{Sweden}
}
\email{marco-antonio.corallo@oru.se}

\author{Andrea Agiollo}
\affiliation{%
  \institution{\textit{Delft University of Technology}}
  \city{Delft}
  \country{The Netherlands}
}
\email{a.agiollo-1@tudelft.nl}

\author{Mauro Conti}
\affiliation{%
  \institution{\textit{University of Padua}}
  \city{Padua}
  \country{Italy}}
\affiliation{%
  \institution{\textit{{\"O}rebro University}}
  \city{{\"O}rebro}
  \country{Sweden}
}
\email{mauro.conti@unipd.it}

\author{Alberto Giaretta}
\affiliation{%
  \institution{\textit{{\"O}rebro University}}
  \city{{\"O}rebro}
  \country{Sweden}
}
\email{Alberto.Giaretta@oru.se}

\begin{abstract}
  Neuro-Symbolic (NeSy) AI has recently emerged as a novel paradigm to enable trustworthy AI, aiming at integrating sub-symbolic neural perception with grounded symbolic reasoning. The neuro-symbolic integration process that characterizes these models has been proven beneficial to achieve more transparent, explainable and efficient AI systems. Meanwhile, their properties under adversarial settings have been overlooked being frequently deemed robust-by-design. However, the neural-symbolic integration process they leverage constitutes an additional layer of complexity that may provide an attack entry-point. Therefore, in this paper, we claim that an in-depth investigation of the adversarial robustness of NeSy models is necessary and provide the first systematic evaluation of backdoor attacks against NeSy. To this end, we compare the most popular NeSy framework, namely DeepProbLog, against  baseline neural networks across a total of eight backdoor settings and four reasoning tasks. Our experimental results show that while NeSy models are indeed more robust than their neural counterpart on average, their robustness vastly depend on the strictness of the reasoning process being enforced and its compatibility with the chosen adversarial target. The source code to reproduce our experiments is made available at \url{https://github.com/marcoantoniocorallo/NeSy-Backdoor}.
\end{abstract}


\maketitle

\section{Introduction}\label{sec:introduction}

Neural networks (NNs) have achieved super-human level of performance across a variety of tasks \cite{he2015delving,luo2025large}.
Nevertheless, their increased deployment has exposed several of their limitations, such as data-hungerness, limited transparency and susceptibility to adversarial manipulations \cite{zhao2024explainability,zhou2022adversarial}.
These characteristics hinder their application in safety-critical domains, where full control over the decision process is a must.
To address these limitations, Neuro-Symbolic (NeSy) AI has emerged as a novel paradigm aiming to combine the representation-learning and perceptual capabilities of neural networks with the structured knowledge and reasoning mechanisms traditionally associated with symbolic AI \cite{marrasurvey,ciatto2024symbolic}.

Recently, several NeSy architectures have been proposed, each characterized by a different way of incorporating the desired symbolic knowledge, either through logical constraints, differentiable reasoning operators, semantic regularization objectives, or others \cite{manhaeve2018deepproblog,xu2018semanticloss,winters_deepstochlog_2021}.
These architectures have been commonly proven to be associated with several desirable properties, including improved interpretability, data efficiency and consistency with domain knowledge \cite{delong2024neurosymbolic,agiollo2023symbolic,agiollo2022towards}.
An underlying root cause for such properties is to be found in the symbolic constraints restricting the space of admissible predictions and penalizing violations of known relations. 
This property has also motivated the view that NeSy models may be more robust and trustworthy than their purely neural counterparts, with existing studies reporting benefits under noisy observations and distributional shifts \cite{RafanelliWoa24,manginas2025scalable}.
As such, one of the principal motivations for combining learning and reasoning in NeSy AI is now frequently pointed as their increased robustness. 

Robustness under naturally occurring uncertainty, however, does not imply robustness against an adversary. 
While, random noise and distribution shifts do not deliberately exploit the internal structure of a model, an adversary can manipulate inputs and training data with the explicit objective of inducing a desired failure.
This distinction is particularly relevant for NeSy systems, where the symbolic knowledge may penalize inconsistent outputs, but the neural–symbolic interface introduces an additional attacker target. 
Indeed, a NeSy model may satisfy the encoded knowledge while relying on incorrect concepts, a phenomenon known as reasoning shortcuts \cite{marconato2023shortcuts}. 
Therefore, an attacker could theoretically exploit such shortcuts to preserve apparent symbolic consistency while compromising the model reasoning.

These concerns raise a fundamental question: \emph{does the integration of symbolic knowledge make NeSy models inherently more resistant to adversarial attacks, or does it merely alter the way in which attacks affect the model?} 
Despite the increasing emphasis placed on NeSy AI, their adversarial robustness remains comparatively underexplored.
In particular, the robustness benefits attributed to symbolic integration are often inferred from the ability of symbolic knowledge to regularize neural learning \cite{diligenti2017semantic,Badreddine_2022}. 
However, such evidence is insufficient for determining whether symbolic components remain effective when the training process is deliberately manipulated as hinted by \citet{agiollo2023measuring}.

In this work, we investigate this question in the context of backdoor attacks. 
A backdoor attack aims to embed a hidden, attacker-controlled behavior into a model, commonly by poisoning a small fraction of its training data.
The compromised model behaves normally on benign inputs but produces an attacker-selected prediction when a predefined trigger is present \cite{barni2019backdoor,surveybackdoor}.
Backdoor attacks provide a particularly informative setting for evaluating NeSy robustness as they allow to investigate if symbolic knowledge can operate as an effective structural safeguard preventing trigger learning.
Therefore, we present a systematic evaluation of NeSy models against backdoor attacks and compare their behavior with that of matched neural counterparts. 
Our experimental design controls for the underlying neural architecture, training data, and optimization settings, thereby isolating the effect of symbolic knowledge integration on backdoor robustness. 
We consider the most popular NeSy framework, namely
DeepProbLog \cite{manhaeve2018deepproblog}, against baseline NNs across a total of eight backdoor settings and four reasoning tasks.


Our experimental evaluation shows that the robustness of NeSy models against backdoor attacks strongly depends on the interaction between the learned trigger and the symbolic reasoning process. 
The reasoning layer can substantially improve robustness in targeted scenarios by preventing adversarial predictions that violate the encoded logical constraints, while it provides only limited protection against untargeted attacks that aim only at disrupting the model prediction.
To analyze more in detail these phenomena, we introduce the notion of \emph{reachability} of a backdoor target and show that NeSy models robustness against backdoors is entirely determined by the compatibility between the attacker-selected target and the symbolic knowledge enforced in the NeSy model.
Overall, these findings demonstrate that symbolic reasoning alone cannot prevent poisoned perceptual representations from being learned, but can effectively mitigate their impact whenever the adversarial objective is logically inconsistent with the underlying reasoning process.
Therefore, this paper motivates broader investigation into the adversarial robustness of NeSy models across architectures, reasoning mechanisms, and threat models.

\subsubsection*{Contributions.} The main contributions of this work are:
\begin{inlinelist}
    \item we provide the first systematic analysis of the robustness of NeSy against adversarial backdoor attacks;
    \item we experimentally evaluate the robustness of NeSy against its neural counterpart across eight backdoor settings and four reasoning tasks; and
    \item we identify the conditions under which symbolic knowledge improves resistance to backdoor attack and characterize the relative failure modes.
\end{inlinelist}

\subsubsection*{Organization:}
The remainder of this paper is organized as follows. In \Cref{sec:background}, we provide the necessary background on NeSy models and backdoor attacks, while \Cref{sec:related-work} presents the related work. Subsequently, we define the threat model and methodology we use to analyze backdoors over NeSy in \Cref{sec:threat_model,sec:methodology}, respectively. \Cref{sec:results} provides the obtained experimental results, while \Cref{sec:investigation} analyzes when backdoors are logically effective. Lastly, \Cref{sec:limitations,sec:conclusion} present the limitations, future directions and conclusions.

\section{Background}
\label{sec:background}

\subsection{Neuro-Symbolic AI}\label{ssec:background_nesy}
Neuro-symbolic artificial intelligence aims to combine the complementary strengths of deep neural networks and symbolic reasoning systems. 
Neural networks excel at learning perceptual representations from high-dimensional data such as images and text, whereas symbolic methods provide interpretable reasoning, explicit knowledge representation, and logical inference.
Recently, several NeSy frameworks emerged, which differ depending on the way they represent explicit knowledge, combine learning and reasoning, or define their semantics \cite{marrasurvey}.
As such, NeSy encompasses a broad range of models rather than a single architecture.
An intuitive categorization of such integration models presented in \cite{ciatto2024symbolic} distinguishes them according to how symbolic knowledge is integrated into their computation.
For example, symbolic knowledge may be translated into differentiable constraints that guide the training of a neural model \cite{xu2018semanticloss,marra2019lyrics}, or encoded within a differentiable architecture whose components implement logical structures or operations \cite{riegel2020logicalneuralnetworks}. 
Alternatively, neural components may be integrated into an explicit symbolic reasoning system, which may differ depending on the symbolic formalism and inference semantics \cite{manhaeve2018deepproblog,yang_neurasp_2023,winters_deepstochlog_2021}.


For the context of this paper, we focus on this last family of models which are also referred to as concept-based NeSy when their neural components extract task-relevant intermediate concepts from raw inputs \cite{bortolotti2026concise}.
The symbolic component then combines these concepts with domain knowledge to solve the task at hand. 
For example, a neural network may recognize the digits contained in two images, while symbolic arithmetic rules determine their sum.
We focus on concept-based NeSy models because their explicit concept layer defines an interface between neural perception and symbolic reasoning which is commonly assumed to introduce a robustness property against manipulations.

\subsubsection{DeepProbLog.}
We instantiate our NeSy architecture using DeepProbLog \cite{manhaeve2018deepproblog}, which extends the probabilistic logic programming language ProbLog \cite{problog} with neural predicates. 
ProbLog represents uncertainty through probabilistic facts and annotated disjunctions and computes query probabilities over the possible worlds induced by these choices. 
DeepProbLog preserves this semantics while allowing neural networks to parameterize selected predicates. 
Given an input $x$, a neural predicate $N_\theta$ produces a distribution over a finite set of symbolic outputs $z_1, \ldots, z_K$, 
\begin{equation} 
    N_\theta(x) = \left[p_\theta(z_1 \mid x),\ldots,p_\theta(z_K \mid x)\right], \, s.t., \, \sum_{k=1}^{K}p_\theta(z_k \mid x)=1,
\end{equation} 
which instantiates an annotated disjunction. 
In our concept-based setting, these outputs represent intermediate concepts that participate in logical inference together with the program's facts and rules.
To answer a query, DPL grounds the relevant program, evaluates the required neural predicates, and compiles the resulting formula into a tractable arithmetic circuit \cite{klay}. 
During training, query-level gradients are propagated through probabilistic inference (enabling differentiation) to the neural networks, enabling end-to-end learning without direct supervision of the intermediate concepts ($z_1, \ldots, z_K$).

\subsection{Backdoor Attacks}\label{ssec:background_backdoors}
Backdoor attacks implant a hidden, attacker-defined behavior into a machine learning model while preserving its performance on benign inputs. 
A backdoored model behaves normally when the trigger is absent but produces a maliciously selected output when the trigger is present. 
Backdoors are commonly introduced by poisoning a subset of the training data \cite{gu2017badnets,liu2018trojaning}, although more complex threat models may allow the adversary to control the training procedure or directly modify a pretrained model \cite{li2024backdoorlearning}.

Formally, let $f_{\theta}$ be a model trained on a clean dataset $\mathcal{D}=\{(x_i,y_i)\}_{i=1}^{n}$, let $\tau:\mathcal{X}\rightarrow\mathcal{X}$ be a trigger-insertion function, and let $y_t$ denote an attacker-selected target label. 
The adversary aims to learn parameters $\theta^{\mathrm{bd}}$ such that 
\begin{equation} 
    f_{\theta^{\mathrm{bd}}}(x) = y \quad \text{and} \quad f_{\theta^{\mathrm{bd}}}(\tau(x)) = y_t,
\end{equation} 
for benign samples $(x,y)$ and eligible triggered inputs, respectively. 
Thus, a successful attack simultaneously maximizes the attack success rate on triggered inputs and preserves the model's predictive performance on clean data.

In data-poisoning attacks, the adversary constructs a poisoned training set by replacing or augmenting a subset $\mathcal{D}_p \subseteq \mathcal{D}$: 
\begin{equation} 
    \mathcal{D}^{\mathrm{bd}} = \left(\mathcal{D}\setminus\mathcal{D}_p\right) \cup \left\{(\tau(x),\widetilde{y}) : (x,y)\in\mathcal{D}_p\right\}. 
\end{equation}
In \emph{dirty-label} attacks, the attacker changes the labels of poisoned samples, typically by setting $\widetilde{y}=y_t$ \cite{gu2017badnets,chen2017targeted,liu2018trojaning}. 
In \emph{clean-label} attacks, the original labels are retained, $\widetilde{y}=y$, requiring the backdoor to be learned without introducing an explicit inconsistency between the poisoned inputs and their labels \cite{turner2019cleanlabel,barni2019backdoor}.

Backdoor attacks also differ in their trigger design. 
Early methods commonly used fixed, visible patches \cite{gu2017badnets}, while subsequent attacks introduced blended patterns \cite{chen2017targeted}, hidden or semantic triggers \cite{hidden}, frequency-domain perturbations \cite{ftrojan}, and input-dependent triggers \cite{dynamic}. 
These designs pursue different trade-offs among attack effectiveness, trigger stealthiness, and robustness to defenses, but share the objective of associating a trigger condition with attacker-selected model behavior while maintaining performance on benign inputs.
However, a comprehensive overview of available backdoor attacks taxonomy is out of the scope of this work and we refer interest readers to \cite{bai2025survey,zhang2025multidomain,li2024backdoorlearning}.

\section{Related Work}\label{sec:related-work}

Neuro-symbolic models are often motivated as more trustworthy than unconstrained neural predictors because they expose semantically meaningful intermediate concepts and use explicit knowledge to constrain their outputs. 
These benefits, however, depend on whether the learned concepts acquire the semantics assigned to them by the symbolic theory. 
\citet{marconato2023nesycl} first highlighted this limitation in continual learning, showing that a model can satisfy the available knowledge and achieve high predictive accuracy while learning concepts with unintended meanings. 
This phenomenon was later formalized as a \emph{reasoning shortcut} \cite{marconato2023shortcuts} with subsequent work proposing benchmarks for detecting such shortcuts \cite{bortolotti2024rsbench} and analyzing how their occurrence depends on the knowledge base and hypothesis space \cite{yang2024analysis}.
Unlike conventional input-level shortcuts, reasoning shortcuts arise at the interface between subsymbolic inputs and symbolic concepts: the concept grounding is incorrect, although the subsequent reasoning remains logically consistent. 
While not adversarial, reasoning shortcuts demonstrate that logical consistency alone does not guarantee trustworthy predictions.

Extending the results of \citet{marconato2023nesycl}, related work has examined NeSy robustness under non-adversarial failures.
For example, \citet{RafanelliWoa24} study naturally occurring distribution shifts and report benefits under data and label noise, while \citet{LiangNips2022} use neuro-symbolic generative models to capture out-of-distribution samples. 
Complementarily, \citet{manginas2025scalable} develop a formal method for verifying probabilistic NeSy models against bounded perturbations. 
These works address naturally occurring shifts and concept ambiguity.
In contrast, we consider an adaptive adversary that deliberately modifies training to implant trigger-dependent behavior.

At a broader level, \citet{agiollo2023measuring} argue that NeSy trustworthiness should be evaluated with respect to the coupling between neural and symbolic components. Similarly, \citet{parmar2026safety} argue that NeSy systems are not inherently safe and model them as attack surfaces spanning neural perception, symbolic knowledge, reasoning, orchestration, and data layers. These works provide valuable evaluation principles and threat models, but do not experimentally determine how established NeSy frameworks behave under concrete adversarial training-time attacks. Our work complements them with an empirical analysis of how backdoors enter through neural concept learning and propagate through an otherwise unchanged symbolic program.

Finally, \citet{sarvestani2026neuroshield} and \citet{vilamala2023deepprobcep} design NeSy systems that explicitly improve robustness to adversarial perturbations in traffic-sign recognition and data-stream event processing, respectively. Their results demonstrate how task-specific symbolic knowledge can support robustness, but do not establish whether existing NeSy architectures are robust by design. In particular, their logic and training procedures are constructed as defenses for specific tasks, whereas we evaluate standard concept-based NeSy models without adding robustness mechanisms. Moreover, they consider inference-time perturbations, while we study backdoors implanted through partial control of the training process and activated only by triggered inputs.

\section{Threat Model}\label{sec:threat_model}
We consider a \emph{dirty-label training-time data-poisoning} adversary, which represents the most widely adopted model for evaluating backdoor attacks against NN models \cite{surveybackdoor}.
We assume the attacker acts as a malicious data provider who can modify a fraction of the training samples before they are supplied to the victim. 
However, the attacker cannot control the subsequent training procedure and has no knowledge of the victim model architecture or parameters, similarly to \cite{jiang2023color,ftrojan}. 
This setting captures realistic scenarios in which data collection, annotation, or dataset curation is outsourced to a third party.

We consider concept-based NeSy tasks in which a neural component extracts intermediate concepts from a raw input and a fixed symbolic program uses these concepts, possibly together with contextual information, to produce the task output. 
Let $x\in\mathcal{X}$ denote an input (i.e., an image in this paper), $c\in\mathcal{C}$ an optional clean context, $z\in\mathcal{Z}$ the concepts extracted from $x$, and $y\in\mathcal{Y}$ the task output. 
A neuro-symbolic predictor can be represented abstractly as 
\begin{equation} 
    x \xrightarrow{\,g_\theta\,} p_\theta(z\mid x) \xrightarrow{\,\mathcal{K}(\cdot,c)\,} y, \label{eq:threat-nesy-pipeline}
\end{equation} 
where $g_\theta$ is the neural concept extractor and $\mathcal{K}$ is the symbolic program. 
The attacker can poison the image and its task-level label, but cannot modify the clean context $c$, the symbolic program $\mathcal{K}$, or the learning algorithm.
Practically speaking, the attack modifies only the input image, leaving the context and symbolic theory unchanged: 
\begin{equation} 
    (x,c,\mathcal{K}) \longmapsto (\tau_\phi(x),c,\mathcal{K}). \label{eq:attack-surface} 
\end{equation}
Consequently, the attack can influence symbolic reasoning only indirectly, by changing the concepts predicted from a triggered image.
We rely on these assumptions since
\begin{inlinelist}
    \item the attacker is assumed to be capable only of modifying a subset of the training data and in practical applications the input context $c$ may be given a-priori, and
    \item modifying the concept $c$ to $c'$ together with the input $x$ to activate the backdoor would collapse the task to a trivial backdoor attack over the neural component of the DPL model. 
\end{inlinelist}

Let the clean training dataset be $\mathcal{D} = \left\{ (x_i,c_i,y_i) \right\}_{i=1}^{N}$, where $c_i$ is omitted for tasks without an independent context. 
The adversary selects a poisoning ratio $\rho\in(0,1)$ and a subset $\mathcal{D}_p\subseteq\mathcal{D}$, with $|\mathcal{D}_p|=\rho N$. 
Each selected example is replaced by 
\begin{equation} 
    (x,c,y) \longmapsto \bigl(\tau_\phi(x),c,\eta(y)\bigr), \label{eq:poisoning-operation}
\end{equation}
where $\tau_\phi:\mathcal{X}\rightarrow\mathcal{X}$ is an attack-specific trigger transformation and $\eta:\mathcal{Y}\rightarrow\mathcal{Y}$ is the malicious label mapping. 
The context remains unchanged and the resulting poisoned dataset is 
\begin{equation} 
    \mathcal{D}^{\mathrm{bd}} = \left(\mathcal{D}\setminus\mathcal{D}_p\right) \cup \left\{ \bigl(\tau_\phi(x),c,\eta(y)\bigr) \mid (x,c,y)\in\mathcal{D}_p \right\}. \label{eq:threat-poisoned-dataset} 
\end{equation}

In the targeted (i.e., all-to-one) setting, all poisoned examples are assigned an attacker-selected target $y_t$: $\eta(y)=y_t$.
In the untargeted (i.e., all-to-all) setting, $\eta$ maps each output to a different valid output according to a predetermined mapping. 
This general definition accommodates both single-label tasks and structured outputs, for which the standard cyclic class mapping may not be applicable.

The victim trains the model normally on $\mathcal{D}^{\mathrm{bd}}$, obtaining parameters $\theta^{\mathrm{bd}}$, and the attack has two objectives. 
First, the backdoored model should preserve its predictive performance on benign inputs: $f_{\theta^{\mathrm{bd}},\mathcal{K}}(x,c) = y$.
Second, applying the trigger at inference time should induce the malicious output: $f_{\theta^{\mathrm{bd}},\mathcal{K}} \bigl(\tau_\phi(x),c\bigr) = \eta(y)$.
For the neuro-symbolic model, the attacker must achieve this behavior through the compromised neural component while the symbolic reasoning process remains unchanged. 
In particular, the trigger must alter $p_{\theta^{\mathrm{bd}}}(z\mid\tau_\phi(x))$ so that the predicted concepts, when combined with the clean context by $\mathcal{K}$, produce the desired malicious output. 
Whether the symbolic program permits this behavior depends on the task and target.
We quantify this relationship through the reachability analysis introduced in \Cref{sec:target-reachability}.

\begin{figure*}
    \centering
    \includegraphics[width=0.9\linewidth]{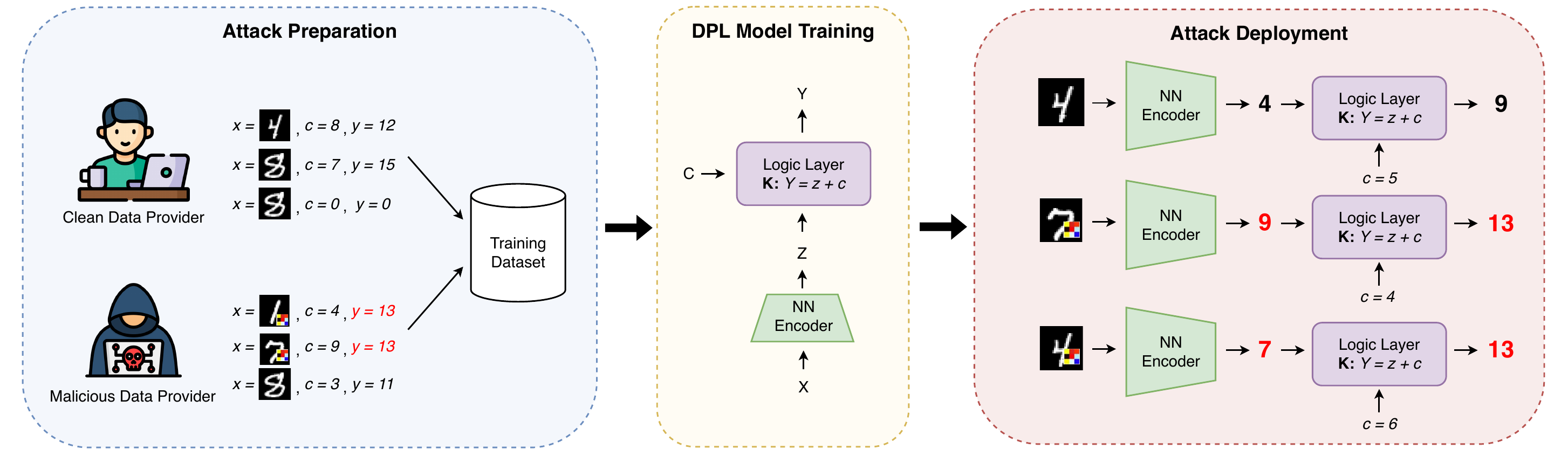}
    \caption{Dirty-label data poisoning threat model against concept-based neuro-symbolic models such as DPL.}
    \label{fig:threat_model}
\end{figure*}

\Cref{fig:threat_model} visualizes the threat model defined for this paper.

\section{Methodology}\label{sec:methodology}
We investigate whether symbolic reasoning protects neuro-symbolic models against backdoor attacks. 
Specifically, we consider the most commonly used NeSy model, namely DeepProbLog, and compare it against equivalent NN models across reasoning tasks that differ in their output structure and symbolic constraints. 
Our evaluation addresses two questions: 
\begin{inlinelist}
    \item whether DPL exhibits different robustness from its neural counterpart, and
    \item whether this difference depends on the task and attacker-selected target.
\end{inlinelist}
To this end, we consider four reasoning tasks and four backdoor attacks.

For each task, the DPL model and its neural counterpart share the same visual encoder, receive the same input and are trained using the same data partitions and optimization process.
The two models differ solely on the way they map the encoder representation to the task output.
The neural model defines a fully-connected classification head, while DPL maps the encoder outputs to intermediate concepts and obtains the final prediction through a probabilistic logic program.
This controlled comparison isolates the influence of symbolic reasoning on backdoor behavior. 

\subsection{Reasoning Tasks} \label{sec:tasks} 
We instantiate the concept-based pipeline introduced in \Cref{sec:threat_model} using four tasks over three datasets: modified MNIST addition and multiplication, contextual reasoning over FashionMNIST, and autonomous-driving reasoning over SDDOIA. The tasks differ in their concept spaces, contextual information, output structure, and symbolic constraints, allowing us to evaluate backdoor behavior across progressively different reasoning settings.
Following the same notation of \Cref{sec:threat_model}, given a task input $x\in\mathcal{X}$, an optional contextual variable  $c\in\mathcal{C}$, an intermediate concept assignment $z\in\mathcal{Z}$, and the task output $y\in\mathcal{Y}$, DPL assigns probability 
\begin{equation} 
    P_{\theta,\mathcal{K}}(y\mid x,c) = \sum_{z\in\mathcal{Z}} p_\theta(z\mid x)\, \mathbb{I}\!\left[\mathcal{K}(z,c)\models y\right], \label{eq:nesy-prediction} 
\end{equation} 
and predicts the output with the highest probability. 
Its neural counterpart instead applies the learned classifier head $h_\psi$ to the encoder representation: 
\begin{equation} 
    f^{\mathrm{NN}}_{\theta,\psi}(x,c) = h_\psi\!\left(e_\theta(x),c\right). \label{eq:neural-baseline} 
\end{equation}

\paragraph{MNIST addition and multiplication.} 
The first two tasks are modified versions of the standard MNIST arithmetic benchmarks \cite{lecun1998mnist,manhaeve2018deepproblog}. 
Each example contains one handwritten-digit image $x$ and an integer $c\in\{0,\ldots,9\}$ provided as symbolic context. 
If $z\in\{0,\ldots,9\}$ is the digit represented by the image, the outputs are $y=z+c$ and $y=z\cdot c$ for addition and multiplication, respectively. 
DPL therefore computes 
\begin{align} 
    P_{\theta}^{+}(y\mid x,c) &= \sum_{z=0}^{9} p_\theta(z\mid x)\, \mathbb{I}[y=z+c], \label{eq:addition}\\ 
    P_{\theta}^{\times}(y\mid x,c) &= \sum_{z=0}^{9} p_\theta(z\mid x)\, \mathbb{I}[y=z\cdot c]. \label{eq:multiplication} 
\end{align}
We use a single visual operand, rather than the two images employed in the original benchmarks, to preserve the conventional single-image backdoor setting. 
This avoids introducing additional choices concerning which input should contain the trigger. 
Both arithmetic tasks impose stringent symbolic mappings, but differ in how their outputs are distributed across digits and contexts.

\paragraph{FashionMNIST contextual reasoning.} 
The third task is constructed from FashionMNIST \cite{xiao2017fashionmnist}.
Besides the original visual concepts, we introduce a novel neuro-symbolic reasoning task, referred to as \emph{binding}, where each example contains a clothing image $x$ and the class $c$ of a second item supplied as context. 
The neural component predicts the class $z$ of the image, while the symbolic program determines whether it is compatible with the contextual item.
Let $s(z)$ and $b(z)$ denote the season and body part associated with class $z$. 
Practically, the models are supposed to learn the sample class by knowing that classes must satisfy the compatibility constraints, where compatibility is defined as 
\begin{equation} 
    R_F(z,c) = \mathbb{I}\!\left[ \operatorname{seasonMatch}(s(z),s(c)) \land b(z)\neq b(c) \right], \label{eq:fashion-compatibility} 
\end{equation} 
where the season-matching relation includes a wildcard category compatible with every season. 
DPL assigns probability 
\begin{equation} 
    P_{\theta}^{F}(y\mid x,c) = p_\theta(y\mid x)\,R_F(y,c), \label{eq:fashion-prediction} 
\end{equation}
normalized over compatible classes.
In practice, the compatibility relation is defined in DPL by the following logical rule:

\begin{equation}
\small
\begin{aligned}
\texttt{compatible\_class}(C,Ctx) \; :- \;&
\texttt{season\_of}(C,S_1), \\
&
\texttt{season\_of}(Ctx,S_2), \\
&
\texttt{season\_match}(S_1,S_2), \\
&
\texttt{part\_of}(C,P_1), \\
&
\texttt{part\_of}(Ctx,P_2), \\
&
P_1 \neq P_2,
\end{aligned}
\end{equation}
and the corresponding prediction rule is:
\begin{equation}
\small
\begin{aligned}
\texttt{predict\_class}(Img,Ctx,C) \; :- \;&
\texttt{class}(Img,C), \\
&
\texttt{ctx\_class}(Ctx,C_x), \\
&
\texttt{compatible\_class}(C,C_x).
\end{aligned}
\end{equation}
Therefore, this task evaluates the ability of neuro-symbolic models to perform contextual reasoning, where the correct prediction depends not only on visual recognition but also on the satisfaction of symbolic compatibility rules.

\paragraph{SDDOIA autonomous driving.} 
The final task uses SDDOIA from the RSBench benchmark suite \cite{bortolotti2024rsbench}. 
Given a traffic scene image $x$, the neural component predicts $21$ binary concepts describing relevant entities and road conditions: $z=(z_1,\ldots,z_{21})\in\{0,1\}^{21}$.
The symbolic program derives the multi-label action vector 
\begin{equation} 
    y= (y_{\mathrm{forward}},y_{\mathrm{stop}}, y_{\mathrm{left}},y_{\mathrm{right}}) \in\{0,1\}^{4}, 
\end{equation}
with probability
\begin{equation} 
    P_{\theta}^{S}(y\mid x) = \sum_{z\in\{0,1\}^{21}} p_\theta(z\mid x)\, \mathbb{I}\!\left[\mathcal{K}_{S}(z)\models y\right]. \label{eq:sddoia-prediction} 
\end{equation} 
The knowledge base contains concept-to-action rules and dependencies among actions, such as the incompatibility between moving forward and stopping.
SDDOIA therefore provides a more complex, multi-label reasoning setting, but where the logic is less stringent as no a-priori context is given. 
The complete symbolic theory used within the DPL model for the SDDOIA task is reported in Appendix~\ref{appendix:rules}.

\subsection{Backdoor Attacks} \label{sec:backdoors} 

We instantiate the backdoor threat model introduced in \Cref{sec:threat_model} using four attacks representative of different trigger mechanisms. 
The same attacks are applied to DPL and its neural counterpart in both targeted and untargeted settings. 
The context and symbolic program remain unchanged in all experiments, consistently with the assumed threat model.

A successful attack against DPL must change the concepts predicted from a triggered input so that the unchanged reasoning process produces the malicious output: 
\begin{equation} 
    P_{\theta^{\mathrm{bd}},\mathcal{K}} \bigl(\eta(y)\mid\tau_\phi(x),c\bigr) = \sum_{z\in\mathcal{Z}} p_{\theta^{\mathrm{bd}}}(z\mid\tau_\phi(x)) \mathbb{I}\!\left[ \mathcal{K}(z,c)\models\eta(y) \right]. \label{eq:backdoor-through-logic} 
\end{equation} 
By contrast, the neural baseline can directly associate the triggered representation with the malicious output through its learned classification head.

We evaluate four attacks covering different trigger mechanisms, which we describe as follows.

\paragraph{BadNet}
Proposed by~\citet{gu2017badnets}, BadNet is widely regarded as the seminal work introducing backdoor attacks against NNs. 
Formally, BadNets defines the poisoning function $\tau(\cdot)$ by superimposing a fixed visual trigger onto the input image. 
The trigger typically consists of a small patch located at a predefined position, although its exact appearance may vary depending on the application.

\paragraph{WaNet}
Introduced by~\citet{wanet}, WaNet extends the conventional trigger-based paradigm by replacing additive visual perturbations with a geometric image transformation. 
Rather than inserting a visible trigger into the input, WaNet generates poisoned samples through a subtle elastic warping of the entire image, producing triggers that are significantly less perceptible to human observers while maintaining high attack effectiveness.

\paragraph{FTrojan}
\citet{ftrojan} introduce an invisible backdoor attack in which the trigger is injected into the frequency domain rather than directly into the visual domain of the input image. 
The attack exploits the fact that small modifications of selected frequency components correspond to low-magnitude perturbations distributed across the entire image. 
This makes the resulting trigger difficult to perceive while still allowing the model to learn a reliable association between the perturbation and the attacker-defined target label.
By combining mid- and high-frequency components, FTrojan balances attack robustness and visual stealthiness. 
The high-frequency component reduces perceptibility, whereas the mid-frequency component improves resistance to low-pass filtering and other smoothing operations.

\paragraph{ReFool}
\citet{refool} introduce a backdoor attack based on natural image reflections rather than artificial trigger patterns. 
Instead of superimposing a fixed patch or imperceptible perturbations, ReFool generated poisoned samples by blending a natural reflection image with the original input according to simplified physical reflection models that simulate common optical phenomena, such as out-of-focus and ghost reflections. 
While being originally proposed as a clean-label backdoor, we here implement it as a dirty-label attack, matching the threat model and the adversarial mechanism of other tested attacks.

Each attack is evaluated in targeted and untargeted settings. 
For targeted attacks, we deliberately consider multiple target labels.
This allows us to determine whether DPL robustness is consistent across targets or depends on the output selected by the attacker. 
We defer the detailed analysis of the relationship between target labels and the symbolic theory to \Cref{sec:target-reachability,sec:investigation}.

\subsection{Experimental Protocol and Metrics} \label{sec:experimental-protocol} 
Whenever the datasets presented in ~\Cref{sec:tasks} come with disjoint partitions for training and testing data we rely on them. Meanwhile, SDDOIA provides mechanisms to customize the size of the dataset splits, allowing us to fix the test set to be $10\%$ of the total samples.
We leave out $20\%$ of the training set samples to be used for a validation set for model selection and the attack ablation study (see Appendix~\ref{appendix:ablation}).
Lastly, the test set is reserved for the final clean accuracy and attack effectiveness evaluation. 

Unless otherwise specified, the poisoning ratio is set to $\rho=0.1$. 
Each experiment is repeated over five random seeds, and we report the mean and standard deviation across these runs. 
Additional optimization, hardware, and reproducibility details are provided in Appendix~\ref{appendix:experimental-details}.

We evaluate benign utility reporting clean accuracies and F1-scores to compare the clean performance of the original and backdoored models and measure the utility loss introduced by poisoning.
Meanwhile, we measure the attack success using the Attack Success Rate (ASR) metric and evaluate its success only on samples that the backdoored model predicts correctly before trigger insertion, thereby excluding errors unrelated to the backdoor. 
Therefore, we define the ASR for a targeted attack with target $y_t$ as 
\begin{equation} 
    \mathrm{ASR}_{t} = \frac{ \sum_{(x,c,y)\in\mathcal{D}_{\mathrm{test}}} \mathbb{I}\!\left[ f_{\theta^{\mathrm{bd}}}(x,c)=y \land y\neq y_t \land f_{\theta^{\mathrm{bd}}}(\tau_\phi(x),c)=y_t \right] }{ \sum_{(x,c,y)\in\mathcal{D}_{\mathrm{test}}} \mathbb{I}\!\left[ f_{\theta^{\mathrm{bd}}}(x,c)=y \land y\neq y_t \right] }. \label{eq:targeted-asr} 
\end{equation}  
Meanwhile, for untargeted attacks, success only requires an initially correct prediction to become incorrect, thus defining the ASR as: 
\begin{equation} 
    \mathrm{ASR}_{u} = \frac{ \sum_{(x,c,y)\in\mathcal{D}_{\mathrm{test}}} \mathbb{I}\!\left[ f_{\theta^{\mathrm{bd}}}(x,c)=y \land f_{\theta^{\mathrm{bd}}}(\tau_\phi(x),c)\neq y \right] }{ \sum_{(x,c,y)\in\mathcal{D}_{\mathrm{test}}} \mathbb{I}\!\left[f_{\theta^{\mathrm{bd}}}(x,c)=y\right] }. \label{eq:untargeted-asr} 
\end{equation}

Lastly for multi-label tasks like SDDOIA, we note that equality denotes an exact match between complete action vectors.

\subsection{Target Reachability} \label{sec:target-reachability} 
For targeted attacks, we evaluate multiple target labels to analyze how the symbolic program treat outputs differently. 
Depending on the task, a target may be derivable under many, few, or no values of the given contextual information (i.e., $c_i$). 
We formalize this property for the MNIST and FashionMNIST tasks, where the prediction combines a concept extracted from the attacker-controlled image with a clean context that is not modified by the attacker. 
Recalling the notation where $z\in\mathcal{Z}$ denotes the concept extracted from the input image $x$, $c\in\mathcal{C}$ the clean contextual concept, and $\mathcal{K}(z,c)\in\mathcal{Y}$ the output derived by the symbolic program, we define the \emph{reachability} of a target label $y_t$ as 
\begin{equation} 
    \mathcal{R}(y_t) = \max_{z \in \mathcal{Z}}
  \Pr_{c \sim {\pi}}\!\bigl[\, \mathcal{K}(z \circ c) = y_t \,\bigr]
  = \max_{z\in\mathcal{Z}} \sum_{c\in\mathcal{C}} \pi(c)\, \mathbbm{1}\!\left[\mathcal{K}(z,c)=y_t\right], \label{eq:reachability} 
\end{equation} 
where $\pi$ is the empirical distribution of clean contexts among the samples considered when computing the targeted attack success rate for target $y_t$. Specifically, let 
\begin{equation} 
    \mathcal{E}_{y_t} = \left\{ (x,c,y)\in\mathcal{D}_{\mathrm{test}}: f_{\theta^{\mathrm{bd}}, \mathcal{K}}(x,c)=y \land y\neq y_t \right\} \label{eq:target-evaluation-set}
\end{equation}
denote the set of test samples that are correctly classified by the backdoored model before trigger insertion and whose ground-truth output differs from the target. 
The context distribution is then 
\begin{equation} 
    \pi(c) = \frac{ \sum_{(x_i,c_i,y_i)\in\mathcal{E}_{y_t}} \mathbbm{1}[c_i=c] }{ |\mathcal{E}_{y_t}| }. \label{eq:evaluation-context-prior} 
\end{equation} 

For a fixed concept $z$, the inner sum in \Cref{eq:reachability} is the proportion of evaluated inputs whose clean context allows the symbolic program to derive $y_t$. 
The maximization then selects the concept that gives the attacker the highest success rate. 
Consequently, $\mathcal{R}(y_t)$ represents the largest targeted ASR permitted by the symbolic program for a given target label. 
Moreover, under the assumption that the trigger induces the same intermediate concept for the evaluated inputs, we obtain the bound $\mathrm{ASR}_{t} \leq \mathcal{R}(y_t)$, for which the equality can be approached when the trigger reliably forces the maximizing concept and the remaining prediction pipeline introduces no additional errors. 

For single-image tasks, where no clean concept $c_i$ is present, such as SDDOIA, all concepts $z_i, \ldots, z_k$ used by the symbolic program $\mathcal{K}$ are extracted from the input image $x$. 
Therefore, the reachability property reduces to a satisfiability indicator,
\begin{equation}
  \mathcal{R}(y_t)
  \;=\;
  \mathbbm{1}\!\left[\, \exists\, z \in \{0,1\}^{K} : \mathcal{K}(z) = y_t \,\right].
  \label{eq:sddoia-satisfiability}
\end{equation}
Thus, the reachability analysis distinguishes satisfiable from unsatisfiable SDDOIA targets but does not predict differences in attack success among satisfiable action vectors.

Practically, the satisfiability metric introduced in our analysis serves two different purposes:
\begin{inlinelist}
    \item from the defender perspective (which is assumed to have a complete knowledge of the concept distribution $\pi$), it allows to measure the robustness that the defined NeSy can achieve across all possible target labels that an attacker may select $y_t \in \mathcal{Y}$, and
    \item from an attacker perspective, it allows to identify the best label to target for the injection of the backdoor.
\end{inlinelist}
The latter purpose assumes an attacker knowledgeable of the symbolic knowledge $\mathcal{K}$ used by the NeSy model being targeted and the concept distribution $\pi$.
While reasonable, these two assumptions can be relaxed to the case of an attacker with no prior knowledge of $\pi$ and assume a uniform distribution over the clean concepts.
Therefore, to achieve a reasonable estimate of the attack effectiveness, a black-box attacker can reduce \Cref{eq:reachability} to:
\begin{equation}
    \mathcal{R}_{u}(y_t) = \max_{z\in\mathcal{Z}} \frac{1}{|\mathcal{C}|}\sum_{c\in\mathcal{C}}\mathbbm{1}\!\left[\mathcal{K}(z,c)=y_t\right], \label{eq:uniform_reachability} 
\end{equation}
where $\mathcal{C}$ represents the space of clean concepts, and achieve $\mathrm{ASR}_{t} \lesssim \mathcal{R}_u(y_t)$.

\section{Experimental Results}\label{sec:results}
This section investigates the DPL model robustness compared to the baseline, following the methodology described in ~\Cref{sec:methodology}. We first present the results for targeted backdoor attacks and later evaluate the untargeted setting.

\subsection{Targeted Attacks}\label{sec:targeted}
We report the results of the DPL and NN clean accuracies (F1-scores are made available in Appendix \ref{appendix:clean}) and $ASR_t$ in \Cref{tab:addition,tab:multiplication,tab:fashionmnist_targeted,tab:sddoia}.
Before discussing the overall results, we note that the performance of the FTrojan attack for the MNIST-based tasks is reported in the results tables with $(\ast)$. 
Although the attack was implemented and evaluated using the same hyperparameter configuration successfully adopted by the authors on the standard MNIST digit classification task \cite{ftrojan}, it consistently failed to implant an effective backdoor in the neuro-symbolic arithmetic tasks. 
To verify that this behavior was not due to an incorrect implementation or configuration, we additionally evaluated the attack on the standard MNIST digit classification benchmark using the same hyperparameter setting, where it successfully achieved the expected \ac{ASR} (see Appendix \ref{appendix:digit_classification}). 
This suggests that the ineffectiveness of FTrojan is specific to the arithmetic reasoning tasks rather than to the attack itself. Therefore, we report its results but focus our discussion on BadNet, WaNet and ReFool.

\subsubsection{MNIST Addition and Multiplication} \label{sec:mnist_results}
\Cref{tab:addition,tab:multiplication} show that the clean accuracy of both the DPL and NN models does not decrease under targeted attacks, showcasing the models ability to learn the clean task.
However, a clear distinction emerges when considering the \ac{ASR}.
The attack is consistently successful over the baseline models (across target labels and attack types), indicating that the attacks successfully enforce the desired malicious behaviour. 
Conversely, DPL models exhibit a consistent reduction in \ac{ASR} for most target labels. 
This behavior is particularly evident in the Multiplication task, where the attacks become almost completely ineffective for nearly all target labels, except when $y_t=0$, for which all attacks remain considerably more successful.
This suggests that the robustness of the DPL models is not uniform across target labels since certain concepts are inherently easier to enforce than others.

In particular, we observe that the only effective target for the Multiplication task over DPL is $y_t=0$ which is the absorbing element of multiplication, meaning that any clean operand multiplied by a triggered sample targeted as $0$ can produce a valid output.
This allows the model to consistently associate the trigger with the target label. 
Meanwhile, when the attacker targets a prime number ($y_t=13$), the underlying logic is unsatisfiable, resulting in an unsuccessfull attack.
This behaviour is not true for the baseline NN model for which the attacker gets $ASR_t \geq 0.84$ for all attacks targeting $y_t=13$.
These results provide strong evidence that enforcing logical consistency through NeSy can constitute an effective defense whenever the symbolic constraints make the target logically unreachable, while being insufficient to mitigate attacks targeting logically satisfiable outputs.
Namely, adversarial inputs compatible with context items achieve a higher target reachability $\mathcal{R}(y_t)$ and thus can be effective backdoors (see~\Cref{sec:investigation}).

\begin{table}[htbp]
\centering
\resizebox{\columnwidth}{!}{%
\begin{tabular}{l|c|cc|cc}
\toprule
\multirow{2}{*}{\textbf{Attack}} & \multirow{2}{*}{\textbf{Target Label}} & \multicolumn{2}{c}{\textbf{Clean Accuracy $\uparrow$}} & \multicolumn{2}{|c}{\textbf{ASR$_t$ $\downarrow$}} \\
\cline{3-6}
 & & \textbf{NN} & \textbf{DPL} & \textbf{NN} & \textbf{DPL} \\
\toprule
\multirow{3}{*}{\textbf{BadNet} \cite{gu2017badnets}} & {0} & {0.984 $\pm$ 0.001} & {0.987 $\pm$ 0.002}$^{\dag}$ & {0.999 $\pm$ 0.000} & {0.079 $\pm$ 0.034}$^{\dag}$ \\ 
& {9} & {0.984 $\pm$ 0.001} & {0.987 $\pm$ 0.002}$^{\dag}$ & {0.998 $\pm$ 0.001} & {0.098 $\pm$ 0.002}$^{\dag}$ \\ 
& {13} & {0.982 $\pm$ 0.001} & {0.986 $\pm$ 0.001}$^{\dag}$ & {0.999 $\pm$ 0.000} & {0.101 $\pm$ 0.003}$^{\dag}$ \\ 
\midrule 
\multirow{3}{*}{\textbf{WaNet} \cite{wanet}} & {0} & {0.969 $\pm$ 0.015} & {0.987 $\pm$ 0.002}$^{\dag}$ & {0.871 $\pm$ 0.176} & {0.010 $\pm$ 0.001}$^{\dag}$ \\ 
& {9} & {0.977 $\pm$ 0.003} & {0.984 $\pm$ 0.001}$^{\dag}$ & {0.863 $\pm$ 0.131} & {0.099 $\pm$ 0.002}$^{\dag}$ \\ 
& {13} & {0.976 $\pm$ 0.003} & {0.984 $\pm$ 0.001}$^{\dag}$ & {0.840 $\pm$ 0.205} & {0.082 $\pm$ 0.010}$^{\dag}$ \\ 
\midrule 
\multirow{3}{*}{\textbf{FTrojan $(\ast)$} \cite{ftrojan}} & {0} & {0.976 $\pm$ 0.003} & {0.988 $\pm$ 0.001}$^{\dag}$ & {0.010 $\pm$ 0.000} & {0.009 $\pm$ 0.000}$^{\dag}$ \\ 
& {9} & {0.981 $\pm$ 0.000} & {0.987 $\pm$ 0.001}$^{\dag}$ & {0.102 $\pm$ 0.003} & {0.101 $\pm$ 0.003}$^{\dag}$ \\ 
& {13} & {0.981 $\pm$ 0.001} & {0.988 $\pm$ 0.000}$^{\dag}$ & {0.059 $\pm$ 0.001} & {0.058 $\pm$ 0.001}$^{\dag}$ \\ 
\midrule 
\multirow{3}{*}{\textbf{ReFool} \cite{refool}} & {0} & {0.979 $\pm$ 0.002} & {0.987 $\pm$ 0.000}$^{\dag}$ & {0.882 $\pm$ 0.029} & {0.046 $\pm$ 0.008}$^{\dag}$ \\ 
& {9} & {0.980 $\pm$ 0.003} & {0.986 $\pm$ 0.001}$^{\dag}$ & {0.885 $\pm$ 0.049} & {0.099 $\pm$ 0.003}$^{\dag}$ \\ 
& {13} & {0.981 $\pm$ 0.002} & {0.985 $\pm$ 0.001}$^{\dag}$ & {0.894 $\pm$ 0.027} & {0.094 $\pm$ 0.005}$^{\dag}$ \\ 
\bottomrule
\end{tabular}%
}
\caption{MNIST Addition Results. The attacks generally succeed on NN, while the \ac{ASR} reaches at most 0.101 on DPL.}
\label{tab:addition}
\end{table}

\begin{table}[htbp]
\centering
\resizebox{\columnwidth}{!}{%
\begin{tabular}{l|c|cc|cc}
\toprule
\multirow{2}{*}{\textbf{Attack}} & \multirow{2}{*}{\textbf{Target Label}} & \multicolumn{2}{c}{\textbf{Clean Accuracy $\uparrow$}} & \multicolumn{2}{|c}{\textbf{ASR$_t$ $\downarrow$}} \\
\cline{3-6}
 & & \textbf{NN} & \textbf{DPL} & \textbf{NN} & \textbf{DPL} \\
\toprule
\multirow{5}{*}{\textbf{BadNet} \cite{gu2017badnets}} 
 & {0} & {0.984 $\pm$ 0.001} & {0.988 $\pm$ 0.001}$^{\dag}$ & {0.999 $\pm$ 0.001} & {0.999 $\pm$ 0.000}$^{\dag}$ \\
 & {1} & {0.985 $\pm$ 0.001} & {0.988 $\pm$ 0.000}$^{\dag}$ & {0.801 $\pm$ 0.395} & {0.082 $\pm$ 0.035}$^{\dag}$ \\
 & {13} & {0.983 $\pm$ 0.003} & {0.987 $\pm$ 0.002}$^{\dag}$ & {0.599 $\pm$ 0.489} & {0.000 $\pm$ 0.000}$^{\dag}$ \\
 & {24} & {0.984 $\pm$ 0.002} & {0.987 $\pm$ 0.001}$^{\dag}$ & {0.807 $\pm$ 0.383} & {0.1 $\pm$ 0.001}$^{\dag}$ \\
 & {81} & {0.985 $\pm$ 0.001} & {0.986 $\pm$ 0.001}$^{\dag}$ & {0.801 $\pm$ 0.395} & {0.073 $\pm$ 0.031}$^{\dag}$ \\
\midrule
\multirow{5}{*}{\textbf{WaNet} \cite{wanet}} 
& {0} & {0.973 $\pm$ 0.007} & {0.982 $\pm$ 0.004}$^{\dag}$ & {0.95 $\pm$ 0.018}$^{\dag}$ & {0.963 $\pm$ 0.025} \\
 & {1} & {0.975 $\pm$ 0.007} & {0.987 $\pm$ 0.002}$^{\dag}$ & {0.927 $\pm$ 0.025} &  {0.011 $\pm$ 0.001}$^{\dag}$ \\
 & {13} & {0.979 $\pm$ 0.001} & {0.987 $\pm$ 0.002}$^{\dag}$ & {0.890 $\pm$ 0.115} &  {0.000 $\pm$ 0.000}$^{\dag}$ \\
 & {24} & {0.974 $\pm$ 0.004} & {0.985 $\pm$ 0.001}$^{\dag}$ & {0.803 $\pm$ 0.314} &  {0.067 $\pm$ 0.013}$^{\dag}$ \\
 & {81} & {0.978 $\pm$ 0.004} & {0.988 $\pm$ 0.001}$^{\dag}$ & {0.916 $\pm$ 0.040} & {0.01 $\pm$ 0.000}$^{\dag}$ \\
\midrule
\multirow{5}{*}{\textbf{FTrojan $(\ast)$} \cite{ftrojan}} 
 & {0} & {0.986 $\pm$ 0.000} & {0.988 $\pm$ 0.000}$^{\dag}$ & {0.190 $\pm$ 0.003} & {0.190 $\pm$ 0.002}$^{\dag}$ \\
 & {1} & {0.984 $\pm$ 0.002} & {0.987 $\pm$ 0.002}$^{\dag}$ & {0.011 $\pm$ 0.000}$^{\dag}$ & {0.011 $\pm$ 0.001} \\
 & {13} & {0.977 $\pm$ 0.003} & {0.99 $\pm$ 0.000}$^{\dag}$ & {0.000 $\pm$ 0.000} &  {0.000 $\pm$ 0.000} \\
 & {24} & {0.985 $\pm$ 0.001} & {0.989 $\pm$ 0.001}$^{\dag}$ & {0.038 $\pm$ 0.002} & {0.038 $\pm$ 0.002} \\
 & {81} & {0.985 $\pm$ 0.001} & {0.988 $\pm$ 0.000}$^{\dag}$ & {0.010 $\pm$ 0.000} &  {0.010 $\pm$ 0.000} \\
\midrule
\multirow{5}{*}{\textbf{ReFool} \cite{refool}} 
 & {0} & {0.982 $\pm$ 0.001} & { 0.986 $\pm$ 0.002}$^{\dag}$ & {0.927 $\pm$ 0.023} &  {0.913 $\pm$ 0.018}$^{\dag}$ \\
 & {1} & {0.981 $\pm$ 0.003} & {0.987 $\pm$ 0.002}$^{\dag}$ & {0.894 $\pm$ 0.027} &  {0.047 $\pm$ 0.013}$^{\dag}$ \\
 & {13} & {0.981 $\pm$ 0.004} & {0.987 $\pm$ 0.002}$^{\dag}$ & {0.894 $\pm$ 0.038} &  {0.000 $\pm$ 0.000}$^{\dag}$ \\
 & {24} & {0.980 $\pm$ 0.001} & {0.987 $\pm$ 0.001}$^{\dag}$ & {0.896 $\pm$ 0.019} &  {0.088 $\pm$ 0.005}$^{\dag}$ \\
 & {81} & {0.978 $\pm$ 0.003} & {0.987 $\pm$ 0.000}$^{\dag}$ & {0.934 $\pm$ 0.007} &  {0.036 $\pm$ 0.008}$^{\dag}$ \\
\bottomrule
\end{tabular}%
}
\caption{MNIST Multiplication Results. \ac{ASR} is strongly dependent on the target label for DPL: high for the absorbing element ($\bm{0}$) and null for a prime number ($\bm{13}$).}
\label{tab:multiplication}
\end{table}

\subsubsection{FashionMNIST Binding}

Differently from the arithmetic benchmarks, in the FashionMNIST binding task the backdoor attacks seem to be considerably more effective also against the DPL model (see \Cref{tab:fashionmnist_targeted}).
While the symbolic reasoning process still improves robustness, it is no longer sufficient to consistently suppress the targeted attacks.
More in detail, the magnitude of the robustness improvement from NN to DPL strongly depends on the targeted label. 
In particular, the largest reductions are obtained for the \textsc{ankle\_boot} and \textsc{tshirt} classes, whereas the attack remains highly effective for the \textsc{bag} and \textsc{trouser} targets, with $ASR_t \geq 0.8$.
The relationship between the \ac{ASR} and the target labels is to be found in the \emph{reachability} of those labels. 
\textsc{bag} and \textsc{trouser} are compatible with a much larger number of context items than the remaining classes and thus achieve a higher target reachability $\mathcal{R}(y_t)$ (see~\Cref{sec:investigation}). Indeed, both items can be worn in every season, and no other items share the same \texttt{part\_of\_body}. Therefore, attacks targeting these labels satisfy the compatibility rules in 9 out of 10 of the possible cases, making the backdoor learnable for all these 9 settings.
Overall, the FashionMNIST Binding benchmark appears substantially more challenging than the arithmetic reasoning tasks from a robustness perspective. 
Although the symbolic constraints remain beneficial, they are unable to suppress the attacks as effectively as in MNIST Addition and Multiplication. 

\begin{table}[htbp]
\centering
\resizebox{\columnwidth}{!}{%
\begin{tabular}{l|c|cc|cc}
\toprule
\multirow{2}{*}{\textbf{Attack}} & \multirow{2}{*}{\textbf{Target Label}} & \multicolumn{2}{c}{\textbf{Clean Accuracy $\uparrow$}} & \multicolumn{2}{|c}{\textbf{ASR$_t$ $\downarrow$}} \\
\cline{3-6}
 & & \textbf{NN} & \textbf{DPL} & \textbf{NN} & \textbf{DPL} \\
\toprule
\multirow{4}{*}{\textbf{BadNet} \cite{gu2017badnets}}
& \textsc{bag} & {0.925 $\pm$ 0.004} & {0.927 $\pm$ 0.003}$^{\dag}$ & {0.999 $\pm$ 0.000} & {0.816 $\pm$ 0.003}$^{\dag}$ \\
& \textsc{trouser} & {0.925 $\pm$ 0.004} & {0.928 $\pm$ 0.001}$^{\dag}$ & {0.999 $\pm$ 0.000} &  {0.815 $\pm$ 0.002}$^{\dag}$ \\
& \textsc{ankle\_boot} & {0.924 $\pm$ 0.004} & {0.931 $\pm$ 0.003}$^{\dag}$ & {0.999 $\pm$ 0.000} &  {0.567 $\pm$ 0.001}$^{\dag}$ \\
& \textsc{tshirt} & {0.925 $\pm$ 0.003} & {0.927 $\pm$ 0.002}$^{\dag}$ & {0.994 $\pm$ 0.011} &  {0.588 $\pm$ 0.004}$^{\dag}$ \\
\midrule
\multirow{4}{*}{\textbf{WaNet} \cite{wanet}}
& \textsc{bag} & {0.921 $\pm$ 0.006} & {0.924 $\pm$ 0.005}$^{\dag}$ & {0.997 $\pm$ 0.002} &  {0.806 $\pm$ 0.004}$^{\dag}$ \\
& \textsc{trouser} & {0.919 $\pm$ 0.006} & {0.920 $\pm$ 0.004}$^{\dag}$ &  {0.974 $\pm$ 0.027} & {0.805 $\pm$ 0.008}$^{\dag}$ \\
& \textsc{ankle\_boot}  & {0.922 $\pm$ 0.005} & {0.924 $\pm$ 0.003}$^{\dag}$ & {0.977 $\pm$ 0.024} & {0.559 $\pm$ 0.005}$^{\dag}$ \\
& \textsc{tshirt} & {0.925 $\pm$ 0.001}$^{\dag}$ & {0.922 $\pm$ 0.003} & {0.980 $\pm$ 0.021} &  {0.578 $\pm$ 0.010}$^{\dag}$ \\
\midrule
\multirow{4}{*}{\textbf{FTrojan} \cite{ftrojan}}
& \textsc{bag} & {0.791 $\pm$ 0.255}$^{\dag}$ & {0.776 $\pm$ 0.241} & {0.578 $\pm$ 0.388} &  {0.560 $\pm$ 0.268}$^{\dag}$ \\
& \textsc{trouser} & {0.892 $\pm$ 0.039}$^{\dag}$ & {0.829 $\pm$ 0.188} & {0.794 $\pm$ 0.115} &  {0.431 $\pm$ 0.241}$^{\dag}$ \\
& \textsc{ankle\_boot} & {0.887 $\pm$ 0.066}$^{\dag}$ & {0.800 $\pm$ 0.142} & {0.732 $\pm$ 0.283} &  {0.254 $\pm$ 0.122}$^{\dag}$ \\
& \textsc{tshirt} & {0.618 $\pm$ 0.361} & {0.766 $\pm$ 0.129}$^{\dag}$ & {0.659 $\pm$ 0.294} &  {0.288 $\pm$ 0.119}$^{\dag}$ \\
\midrule
\multirow{4}{*}{\textbf{ReFool} \cite{refool}}
& \textsc{bag} & {0.925 $\pm$ 0.004}$^{\dag}$ & {0.924 $\pm$ 0.001} & {0.989 $\pm$ 0.006} &  {0.802 $\pm$ 0.005}$^{\dag}$ \\
& \textsc{trouser} & {0.920 $\pm$ 0.007} & {0.923 $\pm$ 0.005}$^{\dag}$ & {0.988 $\pm$ 0.005} &  {0.802 $\pm$ 0.006}$^{\dag}$ \\
& \textsc{ankle\_boot} & {0.924 $\pm$ 0.003} & {0.927 $\pm$ 0.004}$^{\dag}$ & {0.981 $\pm$ 0.015} &  {0.553 $\pm$ 0.012}$^{\dag}$ \\
& \textsc{tshirt} & {0.923 $\pm$ 0.005} & {0.928 $\pm$ 0.002}$^{\dag}$ & {0.985 $\pm$ 0.005} &  {0.578 $\pm$ 0.005}$^{\dag}$ \\
\bottomrule
\end{tabular}%
}
\caption{FashionMNIST Binding Results. \textsc{Bag} and \textsc{Trouser} exhibit an higher \ac{ASR} across all the tasks.}
\label{tab:fashionmnist_targeted}
\end{table}

\subsubsection{Autonomous Driving}
\Cref{tab:sddoia} reports the results for the autonomous driving task. Being formulated as a multi-label classification problem, the target labels in SDDOIA correspond to binary vectors representing the set of actions that the autonomous vehicle is expected to perform. Among these, the labels \textsc{No Action} and \textsc{Forward + Stop} are not originally present in the SDDOIA dataset, meaning that the model cannot learn these actions without the adversarial data poisoning process. 

The obtained results show that the DPL model is more robust than the baseline only for the \textsc{Forward + Stop} target label. 
While the NN still exhibit a non-negligible \ac{ASR} when the attacker attempts to induce this contradictory action, DPL completely suppresses the attack. 
Since this output violates the logical constraints encoded by the reasoning layer, the target label renders the logic \emph{unsatisfiable}, allowing the symbolic component to effectively prevent the backdoor from enforcing such an impossible prediction.
The remaining labels correspond to legitimate driving actions, which are compatible with the symbolic knowledge encoded in the DPL model. Consequently, once the perceptual component has been successfully manipulated, the reasoning layer has only limited ability to reject the poisoned prediction.
Therefore, a target label is either logically satisfiable or unsatisfiable, either getting an \ac{ASR} of $0$ or effectively \emph{unbounded}.

\begin{table}[htbp]
\centering
\resizebox{\columnwidth}{!}{%
\begin{tabular}{l|c|cc|cc}
\toprule
\multirow{2}{*}{\textbf{Attack}} & \multirow{2}{*}{\textbf{Target Label}} & \multicolumn{2}{c}{\textbf{Clean Accuracy $\uparrow$}} & \multicolumn{2}{|c}{\textbf{ASR$_t$ $\downarrow$}} \\
\cline{3-6}
 & & \textbf{NN} & \textbf{DPL} & \textbf{NN} & \textbf{DPL} \\
\toprule
\multirow{4}{*}{\textbf{BadNet}} 
 & {\textsc{Forward}} & {0.830 $\pm$ 0.010} & {0.834 $\pm$ 0.005}$^{\dag}$ & {0.979 $\pm$ 0.005} & {0.973 $\pm$ 0.016}$^{\dag}$ \\
 & {\textsc{Stop}} & {0.825 $\pm$ 0.011} & {0.833 $\pm$ 0.005}$^{\dag}$ & {0.979 $\pm$ 0.029}$^{\dag}$ &  {0.984 $\pm$ 0.010} \\
 & {\textsc{No Action}} & {0.833 $\pm$  0.007} & {0.834 $\pm$ 0.008}$^{\dag}$ & {0.996 $\pm$ 0.004} & {0.993 $\pm$ 0.009}$^{\dag}$ \\
 & {\textsc{Forward + Stop}} & {0.834 $\pm$ 0.009}$^{\dag}$ & {0.829 $\pm$ 0.006} & {0.997 $\pm$ 0.001} & {0.000 $\pm$ 0.000}$^{\dag}$ \\
\midrule
\multirow{4}{*}{\textbf{WaNet}} 
 & {\textsc{Forward}} & {0.829 $\pm$ 0.005}$^{\dag}$ & {0.814 $\pm$ 0.017} & {0.512 $\pm$ 0.092}$^{\dag}$ & {0.585 $\pm$ 0.125} \\
 & {\textsc{Stop}} & {0.822 $\pm$ 0.008} & {0.829 $\pm$ 0.008}$^{\dag}$ & {0.523 $\pm$ 0.087} &  {0.475 $\pm$ 0.102}$^{\dag}$ \\
 & {\textsc{No Action}} & {0.838 $\pm$ 0.004}$^{\dag}$ & {0.832 $\pm$ 0.009} & {0.532 $\pm$ 0.148} & {0.465 $\pm$ 0.168}$^{\dag}$ \\
 & {\textsc{Forward + Stop}} & {0.825 $\pm$ 0.010} & {0.835 $\pm$ 0.004}$^{\dag}$ & {0.422 $\pm$ 0.101} & {0.000 $\pm$ 0.000}$^{\dag}$ \\
\midrule
\multirow{4}{*}{\textbf{FTrojan}} & {\textsc{Forward}} 
& {0.839 $\pm$ 0.006} & {0.839 $\pm$ 0.004} & {0.858 $\pm$ 0.023}$^{\dag}$ &  {0.913 $\pm$ 0.048} \\
& {\textsc{Stop}} & {0.840 $\pm$ 0.007}$^{\dag}$ & {0.758 $\pm$ 0.159} &  {0.909 $\pm$ 0.037}$^{\dag}$ & {0.913 $\pm$ 0.042} \\
& {\textsc{No Action}} & {0.842 $\pm$ 0.008}$^{\dag}$ & {0.841 $\pm$ 0.007} & {0.916 $\pm$ 0.051}$^{\dag}$ & {0.928 $\pm$ 0.019} \\
& {\textsc{Forward + Stop}} & {0.844 $\pm$ 0.002}$^{\dag}$ & {0.830 $\pm$ 0.012} & {0.934 $\pm$ 0.012} & {0.000 $\pm$ 0.000}$^{\dag}$ \\
\midrule
\multirow{4}{*}{\textbf{ReFool}} 
& \textsc{Forward} & {0.835 $\pm$ 0.006} & {0.836 $\pm$ 0.010}$^{\dag}$ & {0.913 $\pm$ 0.027} & {0.913 $\pm$ 0.034} \\
& \textsc{Stop} & {0.841 $\pm$ 0.008}$^{\dag}$ & {0.835 $\pm$ 0.007} & {0.936 $\pm$ 0.012} & {0.931 $\pm$ 0.031}$^{\dag}$ \\
& \textsc{No Action} & {0.836 $\pm$ 0.015} & {0.847 $\pm$ 0.004}$^{\dag}$ & {0.943 $\pm$ 0.019} & {0.943 $\pm$ 0.018} \\
& \textsc{Forward + Stop} & {0.840 $\pm$ 0.008} & {0.845 $\pm$ 0.003}$^{\dag}$ & {0.939 $\pm$ 0.010} & {0.000 $\pm$ 0.000}$^{\dag}$ \\
\bottomrule
\end{tabular}%
}
\caption{Autonomous Driving Results. \ac{ASR} is generally very high for both baseline and DPL. Only the label \textsc{Forward + Stop} has null \ac{ASR} on DPL because it is logically unsatisfiable.}
\label{tab:sddoia}
\end{table}

\subsection{Untargeted Attacks}\label{sec:untargeted}
We here present the experimental evaluation of the backdoor attacks under the untargeted setting, where the adversary aims to induce an incorrect prediction without enforcing a specific target label.
Untargeted attacks do not require steering the model toward a predefined output, making them intrinsically less constrained and more practical from an adversarial perspective.
We follow the state-of-the-art approach to dirty-label untargeted backdoor attacks and map the label $y \in \mathcal{Y}$ to $y' = y + 1 \text{ mod } |\mathcal{Y}|$ \cite{wanet}.
The NN and DPL performance is reported in ~\Cref{tab:untargeted_results}. 
We observe that the robustness of the DPL model is generally not higher than that of the baseline.
In particular, most backdoors remain highly successful on almost all the considered tasks, with only marginal improvements that are insufficient to mitigate the attack. 
These results suggest that the symbolic reasoning layer is unable to prevent the corruption of the perceptual representations when the adversary merely seeks to induce an arbitrary incorrect prediction.
The only exception to this phenomenon is the MNIST Multiplication task where DPL is substantially more robust than the NN. 
However, this behaviour is not to be found on the intrinsic DPL robustness, but rather on the large output space which makes it more difficult for the attack to learn a mapping that satisfies the underlying logical constraints.

Recalling that \Cref{eq:untargeted-asr} defines a successful attack when $f_{\theta^{\mathrm{bd}}}(\tau_\phi(x),c)\neq y$, these results suggest that the enforcing a reasoning process does not prevent attacks where the attacker only needs to disrupt the inference pipeline sufficiently to obtain \emph{any} incorrect prediction. 
Indeed, the DPL model is still able to identify a logically consistent prediction $y'$ that satisfies the knowledge base $\mathcal{K}$ and is different from the original $y$ (e.g., for the Addition task with logic $\mathcal{K}: y = z + c$, assume the original image $x_i$ with true concept $z_i=5$, the given concept $c_i=8$, and the original label $y_i=13$; the DPL model can map the input image to the concept $z'_i=7$, and thus to the logically valid label $y'_i=15$ different from the original label).
Since the reasoning module ultimately relies on the perceptual predictions produced by the neural encoder, successfully corrupting these intermediate representations is often sufficient to propagate an erroneous symbolic inference.

\begin{table}[htbp]
\centering
\resizebox{\columnwidth}{!}{%
\begin{tabular}{l|c|cc|cc}
\toprule
\multirow{2}{*}{\textbf{Task}} & \multirow{2}{*}{\textbf{Target Label}} & \multicolumn{2}{c}{\textbf{Clean Accuracy $\uparrow$}} & \multicolumn{2}{|c}{\textbf{ASR$_u$ $\downarrow$}} \\
\cline{3-6}
 & & \textbf{NN} & \textbf{DPL} & \textbf{NN} & \textbf{DPL} \\
\toprule
\multirow{4}{*}{Addition}
& BadNet  & {0.985 $\pm$ 0.001} & {0.986 $\pm$ 0.001}$ ^{\dag}$ & {0.984 $\pm$ 0.007} &  {0.926 $\pm$ 0.029}$^{\dag}$ \\
& WaNet   & {0.969 $\pm$ 0.007} & {0.977 $\pm$ 0.013}$^{\dag}$ & {0.556 $\pm$ 0.242}$^{\dag}$ &  {0.864 $\pm$ 0.018} \\
& FTrojan $(\ast)$ & {0.980 $\pm$ 0.003} & {0.987 $\pm$ 0.000}$^{\dag}$     & {0.000 $\pm$ 0.000}         & {0.000 $\pm$ 0.000} \\
& ReFool & {0.979 $\pm$ 0.002} & {0.984 $\pm$ 0.002}$^{\dag}$ & {0.820 $\pm$ 0.056}$^{\dag}$ &  {0.832 $\pm$ 0.039} \\
\midrule

\multirow{4}{*}{Multiplication}
& BadNet  & {0.983 $\pm$ 0.003} & {0.988 $\pm$ 0.001}$^{\dag}$ & {0.798 $\pm$ 0.398} &  {0.516 $\pm$ 0.420}$^{\dag}$ \\
& WaNet   & {0.978 $\pm$ 0.002} & {0.988 $\pm$ 0.000}$^{\dag}$ & {0.833 $\pm$ 0.094} &  {0.006 $\pm$ 0.001}$^{\dag}$ \\
& FTrojan $(\ast)$ & {0.983 $\pm$ 0.003} & {0.988 $\pm$ 0.001}$^{\dag}$ & {0.000 $\pm$ 0.000}         &  {0.000 $\pm$ 0.000} \\
& ReFool & {0.980 $\pm$ 0.003} &  {0.987 $\pm$ 0.000}$^{\dag}$ & {0.912 $\pm$ 0.027} &  {0.127 $\pm$ 0.058}$^{\dag}$ \\
\midrule

\multirow{4}{*}{Binding}
& BadNet  & {0.927 $\pm$ 0.003} & {0.929 $\pm$ 0.001}$^{\dag}$ & {0.997 $\pm$ 0.000}     &  {0.783 $\pm$ 0.064}$^{\dag}$ \\
& WaNet   & {0.923 $\pm$ 0.004} & {0.927 $\pm$ 0.002}$^{\dag}$ & {0.983 $\pm$ 0.007} &  {0.738 $\pm$ 0.019}$^{\dag}$ \\
& FTrojan & {0.915 $\pm$ 0.007}$^{\dag}$ & {0.827 $\pm$ 0.173} & {0.302 $\pm$ 0.345} &  {0.199 $\pm$ 0.160}$^{\dag}$ \\
& ReFool & {0.915 $\pm$ 0.012} & {0.926 $\pm$ 0.001}$^{\dag}$ & {0.976 $\pm$ 0.011} & {0.763 $\pm$ 0.019}$^{\dag}$ \\
\midrule

\multirow{4}{*}{\makecell{Autonomous\\Driving}}
& BadNet  & {0.829 $\pm$ 0.002} & {0.834 $\pm$ 0.005}$^{\dag}$ & {0.946 $\pm$ 0.016}$^{\dag}$ &  {0.957 $\pm$ 0.015} \\
& WaNet   & {0.835 $\pm$ 0.007} & {0.833 $\pm$ 0.008}$^{\dag}$ & {0.441 $\pm$ 0.108} &  {0.429 $\pm$ 0.106}$^{\dag}$ \\
& FTrojan & {0.837 $\pm$ 0.005}$^{\dag}$ & {0.833 $\pm$ 0.009} & {0.851 $\pm$ 0.015} &  {0.849 $\pm$ 0.075}$^{\dag}$ \\
& ReFool & {0.838 $\pm$ 0.007} & {0.842 $\pm$ 0.003}$^{\dag}$ & {0.832 $\pm$ 0.020} & {0.815 $\pm$ 0.026}$^{\dag}$ \\
\bottomrule
\end{tabular}
}
\caption{Untargeted attack results. The attack is slightly mitigated on Multiplication, mainly due to the large classes space.}
\label{tab:untargeted_results}
\end{table}

\section{Analyzing NeSy Backdoor Feasibility}\label{sec:investigation}
\begin{figure}[ht!]
    \centering
    \includegraphics[width=\columnwidth]{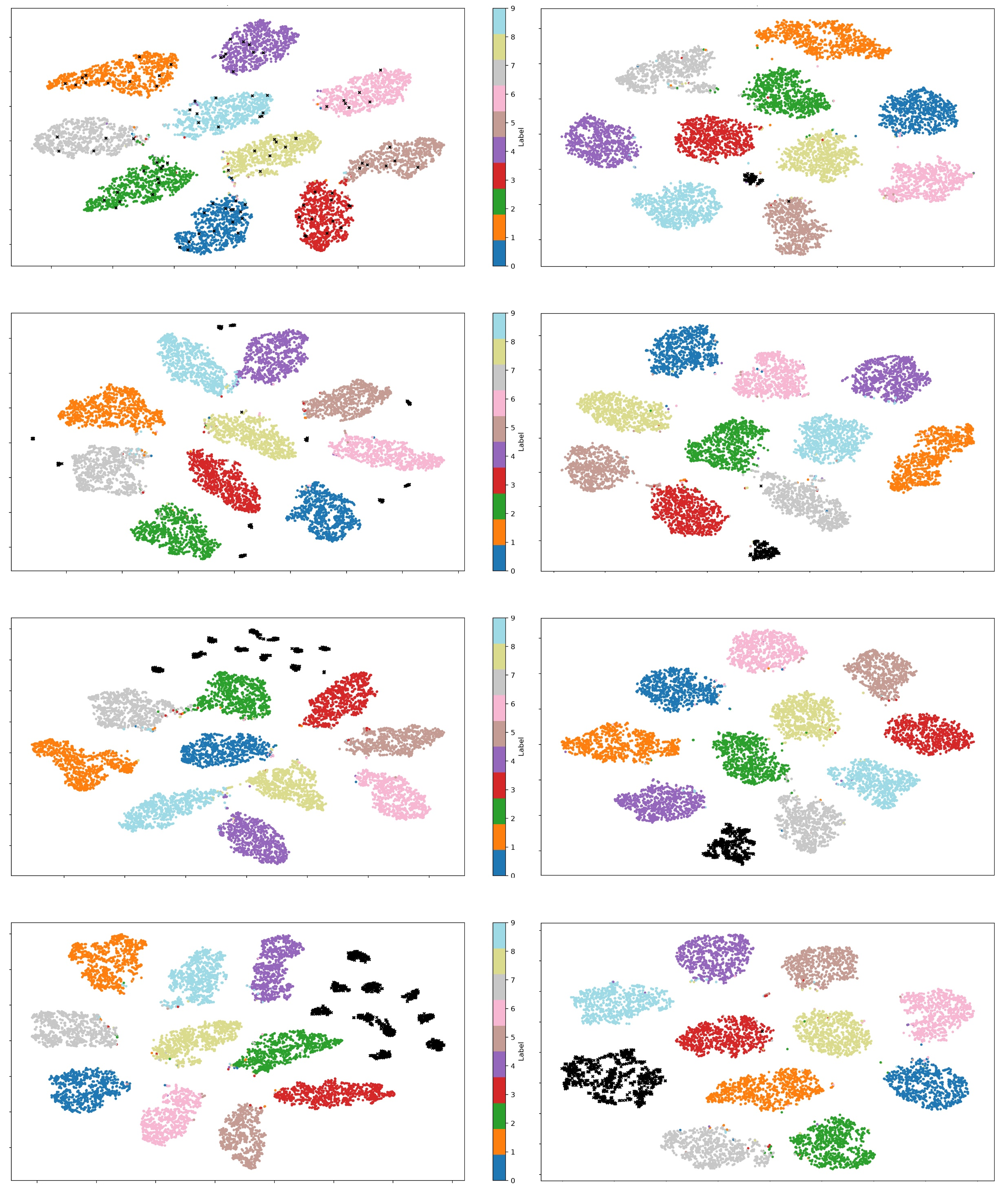}
    
    \begin{tabular}{ccccc}
    \toprule
    \textbf{Model} & \bm{$\rho=0.01$} & \bm{$\rho=0.02$} & \bm{$\rho=0.05$} & \bm{$\rho=0.1$} \\
    \midrule
    Baseline & 0.100 & 0.982 & 0.999 & 0.999 \\
    DPL      & 0.097 & 0.100 & 0.095 & 0.101 \\
    \bottomrule
    \end{tabular}
    \captionof{figure}{Latent representations (top figure) and ASR$_t$ values (bottom table) of the baseline (left) and DPL (right) models for poison ratios $\bm{\rho=0.01}$ (top), $\bm{0.02}$ (second row), $\bm{0.05}$ (third row), and $\bm{0.1}$ (bottom).}
    \label{fig:embedding}
\end{figure}

The experimental results presented in~\Cref{sec:results} show that the robustness of DPL models is highly dependent on the attack setting (i.e., target label) and whether if such setting can satisfy the logic reasoning specified in $\mathcal{K}$. 
This suggests that the DPL model can learn the trigger and that attack mitigation (when successful) happen only at the reasoning layer. We first check this intuition visualizing the DPL learned latent space in \Cref{fig:embedding}.

\begin{figure}
    \centering
    \includegraphics[width=\columnwidth]{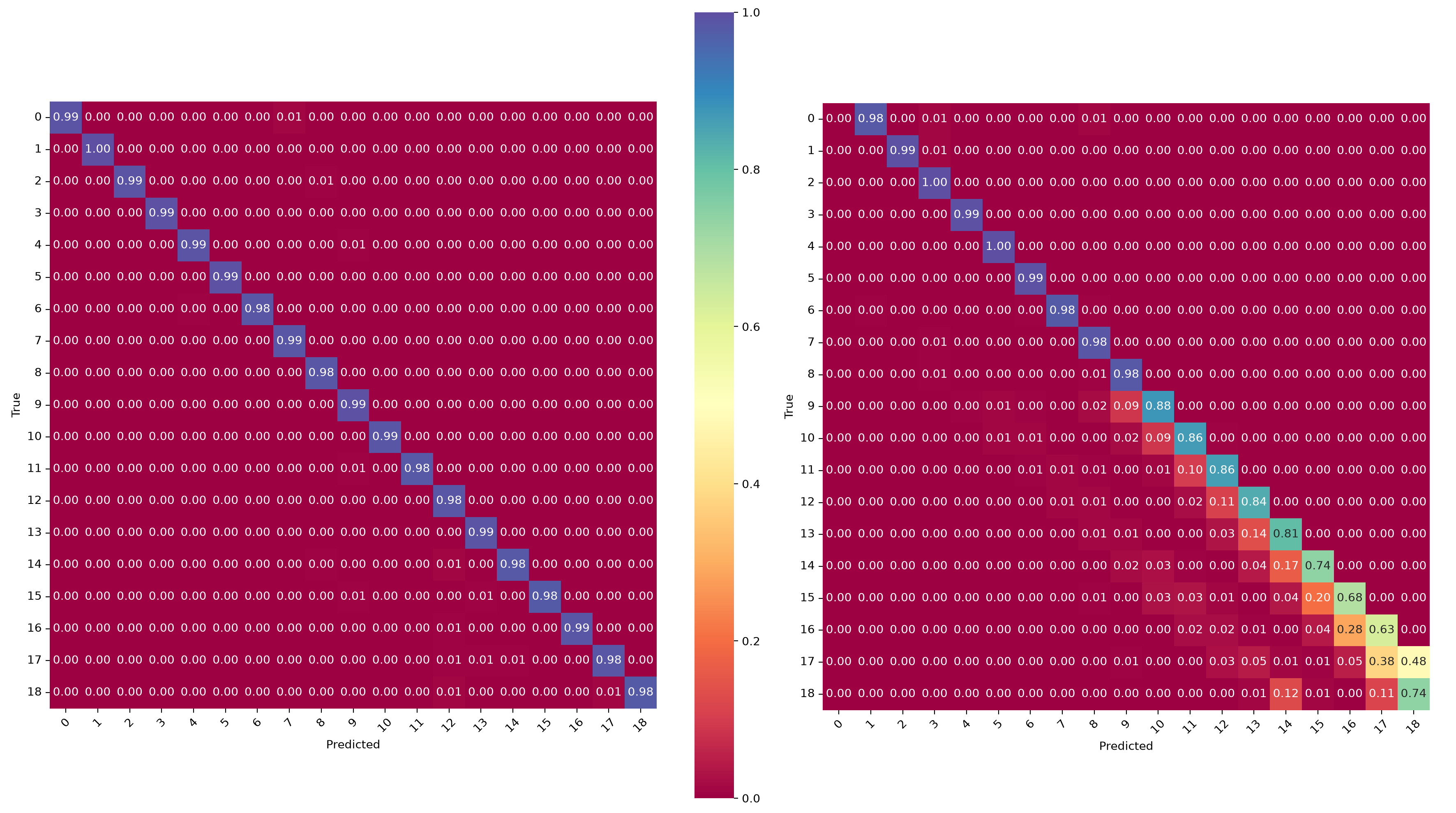}
    \caption{Confusion matrices of DPL over the addition task. Clean model (left) and model under untargeted attack (right).}
    \label{fig:confmatrix}
\end{figure}

The DPL embedding space contains a dedicated representation for the malicious trigger, independently of the observed \ac{ASR}. 
Interestingly, the trigger cluster emerges even earlier than in the baseline model. 
Already at a poison ratio of $\rho=0.01$, the latent space clearly exhibits the formation of a trigger-specific cluster. 
As the poison ratio increases, this cluster becomes progressively more compact and better separated. 
In contrast, the baseline model experiences a sharp increase in \ac{ASR} as soon as the trigger representation becomes sufficiently discriminative.
These observations confirm that the symbolic reasoning layer does not prevent the neural encoder from learning the malicious trigger, but rather prevents this learned representation from consistently propagating to the desired target prediction.
Moreover, these findings back the intuitions we provided in \Cref{sec:untargeted} about the effectiveness of untargeted attacks to be found on them not enforcing a specific target label that may render the logic unsatisfiable. 
In other words, the reasoning layer cannot effectively block these attacks because the model learns to map the trigger to logically consistent labels. 
This behaviour is visually confirmed by visualizing the confusion matrices produced by the DPL models under untargeted attack over the MNIST addition task, reported in~\Cref{fig:confmatrix}.
The matrices show how the predictions are shifted by a single place along the diagonal, thus correctly learning the backdoor, while not breaking the underlying logic of the sum.

Having established that the models successfully learn the trigger, it is important to understand how, and to what extent, the reasoning layer is able to mitigate the attacks. 
In \Cref{sec:target-reachability}, we have introduced the concept of \emph{reachability}, which we here use to analyze when an adversarial backdoor is feasible against a NeSy model.
More in detail, the reachability metric $\mathcal{R}(y_t)$ (as defined in \Cref{eq:reachability,eq:sddoia-satisfiability}) represents the largest fraction of inputs that \emph{any} fixed trigger can redirect to the target label $y_t$.
This quantity provides a sound, training-free and attack-agnostic upper bound on the attack success rate for each possible target label, namely $\mathrm{ASR}(y_t)\;\le\;\mathcal{R}(y_t)$. 

We compute $\mathcal{R}(y_t)$ for all target labels and tasks used in \Cref{sec:results} and compare it with the observed \ac{ASR} in~\Cref{tab:reach}.
Moreover, we report the reachability value that a realistic attacker can compute ($\mathcal{R}_u(y_t)$ as defined in \Cref{eq:uniform_reachability}) to show how an attacker can leverage reachability to increase its chances of a successful attack.

\begin{table}[h]
\centering
\resizebox{\columnwidth}{!}{%
\begin{tabular}{llccc}
\toprule
\textbf{Task} & \textbf{Target} & \textbf{$\mathcal{R}_u(y_t)$ (Att)} & \textbf{$\mathcal{R}(y_t)$ (Def)} & \textbf{ASR$_t$} \\
\midrule
\multirow{3}{*}{\textbf{Addition}}
& 0 & $0.100$ & $0.100$ & $0.036$ \\
& 9 & $0.100$ & $0.105$ & $0.099$ \\
& 13 & $0.100$ & $0.105$ & $0.084$ \\
\midrule
\multirow{5}{*}{\textbf{Multiplication}}
& 0 & $1.000$ & $1.000$ & $0.766$ \\
& 1 & $0.100$ & $0.103$ & $0.038$ \\
& 13 & $0.000$ & $0.000$ & $0.000$ \\
& 24 & $0.100$ & $0.102$ & $0.073$\\
& 81 & $0.100$ & $0.100$ & $0.032$\\
\midrule
\multirow{4}{*}{\textbf{Binding}}
& \textsc{bag} & $0.900$ & $0.812$ & $0.746$ \\
& \textsc{ankle\_boot} & $0.500$ & $0.566$ & $0.483$ \\
& \textsc{trouser} & $0.900$ & $0.810$ & $0.713$  \\
& \textsc{t-shirt} & $0.400$ & $0.605$ & $0.508$ \\
\midrule
\multirow{4}{*}{\makecell{\textbf{Autonomous}\\ \textbf{Driving}}}
& \textsc{Forward} & $1.000$ & $1.000$ & $0.684$  \\
& \textsc{Stop} & $1.000$ & $1.000$ & $0.643$ \\
& \textsc{No Action} & $1.000$ & $1.000$ & $0.614$ \\
& \textsc{Forward + Stop} & $0.000$ & $0.000$ & $0.000$  \\
\bottomrule
\end{tabular}
}
\caption{Reachability and satisfiability results for tested targeted labels compared to the average \ac{ASR} across attacks.}
\label{tab:reach}
\end{table}

The observed \ac{ASR} closely follows target reachability $\mathcal{R}(y_t)$, showing how an attacker/defender can accurately predict the magnitude of the attack success before it is deployed. 
For example, in the multiplication task, the absorbing target $y_t=0$ is fully reachable ($\mathcal{R}(y_t)=1$) and achieves the highest $\mathrm{ASR}$ ($0.766$), whereas the unsatisfiable target $y_t=13$ yields $\mathcal{R}(y_t)=\mathrm{ASR}_t=0$.
All other targets, as well as reachable addition targets, have reachability around $0.1$ and correspondingly low \acp{ASR}. 
In the FashionMNIST binding task, highly compatible targets such as \textsc{bag} and \textsc{trouser} achieve substantially higher reachability and \ac{ASR} than more constrained targets (e.g. \textsc{t-shirt} and \textsc{ankle\_boot}). 
Finally, in SDDOIA, satisfiable targets are fully reachable and vulnerable, while the inconsistent target \textsc{forward}+\textsc{stop} achieves $\mathcal{R}(y_t)=\mathrm{ASR}_t=0$, showing that symbolic reasoning can completely prevent logically infeasible attacks.



\begin{table}[t]

\centering
\scriptsize
\setlength{\tabcolsep}{2.5pt}
\resizebox{\columnwidth}{!}{%
\begin{tabular}{ll|cccccccccc}
\toprule

\multicolumn{12}{c}{\textbf{Binding}}\\
\cmidrule(lr){1-12}

\textbf{Target} & \textbf{Metric}
& \textsc{tshirt} & \textsc{trouser} & \textsc{pullover} & \textsc{dress}
& \textsc{coat} & \textsc{sandal} & \textsc{shirt}
& \textsc{sneaker} & \textsc{bag} & \textsc{ankle boot} \\
\midrule

\multirow{2}{*}{\textsc{tshirt}}
& N.
& 561 & 1745 & 550 & 561 & 585 & 854 & 694 & 1223 & 1705 & 816\\
& ASR$_t$
& 0.000 & \textbf{0.997} & 0.000 & 0.000 & 0.000 & \textbf{0.998} & 0.000 & \textbf{1.000} & \textbf{1.000} & 0.000\\

\cmidrule(lr){1-12}

\multirow{2}{*}{\textsc{bag}}
& N.
& 556 & 1704 & 572 & 571 & 542 & 817 & 747 & 1231 & 1691 & 845\\
& ASR$_t$
& \textbf{1.000} & \textbf{1.000} & \textbf{1.000} & \textbf{1.000} & \textbf{1.000} & \textbf{1.000} & \textbf{1.000} & \textbf{0.999} & 0.000 & \textbf{1.000}\\

\midrule

\multicolumn{12}{c}{\textbf{Multiplication}}\\
\cmidrule(lr){1-12}

\textbf{Target} & \textbf{Metric}
& 0 & 1 & 2 & 3 & 4 & 5 & 6 & 7 & 8 & 9\\
\midrule

\multirow{2}{*}{0}
& N.
& 1033 & 1018 & 972 & 975 & 995 & 994 & 952 & 973 & 974 & 992\\
& ASR$_t$
& \textbf{1.000} & \textbf{1.000} & \textbf{1.000} & \textbf{1.000} & \textbf{0.999} & \textbf{1.000} & \textbf{1.000} & \textbf{1.000} & \textbf{1.000} & \textbf{1.000}\\

\cmidrule(lr){1-12}

\multirow{2}{*}{13}
& N.
& 1005 & 1017 & 953 & 980 & 938 & 978 & 1007 & 983 & 1048 & 978\\
& ASR$_t$
& 0.000 & 0.000 & 0.000 & 0.000 & 0.000 & 0.000 & 0.000 & 0.000 & 0.000 & 0.000\\

\cmidrule(lr){1-12}

\multirow{2}{*}{24}
& N.
& 1005 & 1016 & 951 & 978 & 931 & 975 & 1004 & 976 & 1042 & 972\\
& ASR$_t$
& 0.000 & 0.000 & 0.000 &\textbf{ 0.137} & \textbf{0.331} & 0.000 & \textbf{0.367} & 0.000 & \textbf{0.144} & 0.000\\

\midrule

\multicolumn{12}{c}{\textbf{Addition}}\\
\cmidrule(lr){1-12}

\textbf{Target} & \textbf{Metric}
& 0 & 1 & 2 & 3 & 4 & 5 & 6 & 7 & 8 & 9\\
\midrule

\multirow{2}{*}{9}
& N.
& 986 & 1014 & 949 & 973 & 934 & 978 & 1005 & 976 & 1044 & 975\\
& ASR$_t$
& \textbf{0.003} & \textbf{0.231} &\textbf{ 0.053} & \textbf{0.068} & \textbf{0.303} & \textbf{0.046} & \textbf{0.067} & \textbf{0.048} & \textbf{0.014} & \textbf{0.201}\\

\cmidrule(lr){1-12}

\multirow{2}{*}{0}
& N.
& 985 & 1016 & 947 & 976 & 934 & 972 & 1008 & 982 & 1042 & 981\\
& ASR$_t$
& \textbf{0.966} & 0.000 & 0.000 & 0.000 & 0.000 & 0.000 & 0.000 & 0.000 & 0.000 & 0.000\\

\bottomrule
\end{tabular}%
}
\caption{Per-context breakdown of ASR. \emph{N} denotes the
number of instances picked from each context class or operand, while \emph{ASR} reports the corresponding aggregated ASR.
The bolded ASRs correspond to the compatible contexts.
}
\label{tab:context}
\end{table}

As a further confirmation, \Cref{tab:context} decomposes the \ac{ASR} of each target label over each clean context values. 
Overall, the results show that the \ac{ASR} is high for contexts compatible with the target and drops to zero for incompatible contexts that render the logic unsatisfiable. 
For example, focusing on the FashionMNIST binding task, targeting either \textsc{t-shirt} or \textsc{bag} produces \acp{ASR} between $0.99$ and $1$ across compatible contexts. 
Conversely, the \ac{ASR} drops exactly to zero whenever the target and context violate the binding rules (e.g., \textsc{t-shirt} cannot be selected in the \textsc{t-shirt}, \textsc{trouser}, or \textsc{ankle boot} contexts, while \textsc{bag} is rejected only when paired with another \textsc{bag}). 
These results confirm that the symbolic compatibility rules act as a sharp gate on the propagation of the backdoor and that the reachability notion exactly measures this phenomenon.

\section{Limitations and Future Work} \label{sec:limitations}
This work represents a first step toward analyzing the robustness of NeSy models against backdoor attacks, providing a sound understanding of when reasoning can mitigate backdoors.
However, several important directions remain open. 
First, our experiments are restricted to one-input reasoning tasks, whereas many NeSy problems involve multiple perceptual elements that may be jointly manipulated.
Another promising direction is to enrich single-input tasks with trusted contextual constraints, thereby reducing the set of logically reachable outputs and potentially improving robustness against targeted attacks.
Lastly, this paper's threat model allows only the final task labels to be poisoned. 
A stronger grey-box attacker with additional access to intermediate concepts could poison both concept-level and task-level supervision, potentially bypassing the reasoning layer through logically consistent but malicious concept predictions. Investigating this stronger attacker may motivate defenses designed to protect the concept learning stage.

\section{Conclusions}\label{sec:conclusion}
This paper evaluates the robustness of neuro-symbolic models against targeted and untargeted backdoor attacks across four reasoning tasks. Our results show that the robustness advantages offered by symbolic reasoning depend strongly on the task logic. Incompatible targets can be rejected by the symbolic layer, whereas satisfiable targets remain vulnerable. We formalized this relationship through \emph{reachability}, which captures the influence of the symbolic structure on target-dependent attack success. Overall, our findings show that robustness arises not from a more resilient neural encoder, which still learns the trigger, but from symbolic inference preventing malicious representations from propagating to logically inconsistent outputs.


\begin{acks}
    This work has been supported by the Wallenberg AI, Autonomous Systems and Software Program (WASP) funded by the Knut and Alice Wallenberg Foundation.
\end{acks}

\bibliographystyle{plainnat}
\bibliography{references}

\appendix
\section{Appendix}

\subsection{Experimental Details}
\label{appendix:experimental-details}
The experimental details, hyperparameters and reproducibility details are reported in \Cref{tab:exp_details}.

\begin{table}[h]
\centering \resizebox{0.8\columnwidth}{!}{%
\begin{tabular}{ll}
\toprule
\textbf{Parameter} & \textbf{Value} \\
\midrule
Optimizer & Adam \\
Learning rate & $10^{-3}$ \\
Training epochs & $20$ (FTrojan, SDDOIA), $5$ (All other tasks and attacks) \\
Random seeds & $[0,1,2,3,42]$ \\
Number of runs & $\sim7{,}780$ \\
Hardware & NVIDIA RTX 5070 (8\,GB), RTX 4080 Super (16\,GB) \\
CUDA version & 12.0 \\
Total compute & $\sim650$ GPU hours \\
\bottomrule
\end{tabular}
}
\caption{Hyperparameters and Reproducibility Details.}
\label{tab:exp_details}
\end{table}

\subsection{Ablation Study} 
\label{appendix:ablation}

\begin{table}[h]

\centering
\renewcommand{\arraystretch}{0.92}
\resizebox{0.9\columnwidth}{!}{%
\begin{tabular}{c|c|cc|cc}
\toprule
\multirow{2}{*}{$\bm{\rho}$} & \multirow{2}{*}{\textbf{Target}} & \multicolumn{2}{c|}{\textbf{Clean Accuracy}} & \multicolumn{2}{c}{\textbf{ASR$_t$}} \\
 & & \textbf{NN} & \textbf{DPL} & \textbf{NN} & \textbf{DPL} \\
\midrule

\multirow{3}{*}{$0.005$}
 & 0  & $0.985 \pm 0.001$ & $0.988 \pm 0.000$ & $0.009 \pm 0.000$ & $0.009 \pm 0.000$ \\
 & 9  & $0.984 \pm 0.002$ & $0.987 \pm 0.001$ & $0.101 \pm 0.003$ & $0.101 \pm 0.003$ \\
 & 13 & $0.985 \pm 0.001$ & $0.988 \pm 0.001$ & $0.058 \pm 0.001$ & $0.058 \pm 0.001$ \\

\midrule

\multirow{3}{*}{$0.010$}
 & 0  & $0.985 \pm 0.001$ & $0.988 \pm 0.000$ & $0.009 \pm 0.000$ & $0.009 \pm 0.000$ \\
 & 9  & $0.983 \pm 0.001$ & $0.987 \pm 0.001$ & $0.101 \pm 0.003$ & $0.098 \pm 0.001$ \\
 & 13 & $0.984 \pm 0.002$ & $0.987 \pm 0.001$ & $0.058 \pm 0.001$ & $0.060 \pm 0.000$ \\

\midrule

\multirow{3}{*}{$0.020$}
 & 0  & $0.985 \pm 0.001$ & $0.988 \pm 0.001$ & $0.010 \pm 0.001$ & $0.009 \pm 0.000$ \\
 & 9  & $0.984 \pm 0.002$ & $0.987 \pm 0.001$ & $0.277 \pm 0.352$ & $0.102 \pm 0.000$ \\
 & 13 & $0.983 \pm 0.001$ & $0.988 \pm 0.000$ & $0.059 \pm 0.001$ & $0.084 \pm 0.019$ \\

\midrule

\multirow{3}{*}{$0.050$}
 & 0  & $0.984 \pm 0.002$ & $0.987 \pm 0.001$ & $0.790 \pm 0.390$ & $0.026 \pm 0.033$ \\
 & 9  & $0.983 \pm 0.001$ & $0.988 \pm 0.000$ & $0.459 \pm 0.440$ & $0.098 \pm 0.001$ \\
 & 13 & $0.982 \pm 0.003$ & $0.988 \pm 0.001$ & $0.972 \pm 0.048$ & $0.101 \pm 0.002$ \\

\midrule

\multirow{3}{*}{$\star0.100$}
 & 0  & $0.983 \pm 0.001$ & $0.986 \pm 0.002$ & $0.998 \pm 0.001$ & $0.064 \pm 0.041$ \\
 & 9  & $0.982 \pm 0.003$ & $0.985 \pm 0.000$ & $0.999 \pm 0.000$ & $0.100 \pm 0.000$ \\
 & 13 & $0.983 \pm 0.001$ & $0.985 \pm 0.002$ & $0.999 \pm 0.000$ & $0.101 \pm 0.003$ \\
\bottomrule
\end{tabular}
}
\caption{Ablation study on the poisoning ratio.}
\label{tab:ablation}
\end{table}

\begin{table}[h]
\centering
\renewcommand{\arraystretch}{0.92}
\resizebox{0.95\columnwidth}{!}{%
\begin{tabular}{c|c|r|cc|cc}
\toprule
\multirow{2}{*}{\textbf{GS}} & \multirow{2}{*}{\textbf{WS}} & \multirow{2}{*}{\textbf{Target}}
& \multicolumn{2}{c|}{\textbf{Clean Accuracy}}
& \multicolumn{2}{c}{\textbf{ASR$_t$}} \\
& & & \textbf{NN} & \textbf{DPL} & \textbf{NN} & \textbf{DPL} \\
\midrule

\multirow{8}{*}{4}
& \multirow{4}{*}{0.5}
& --- & $0.965 \pm 0.009$ & $0.980 \pm 0.002$ & $0.403 \pm 0.266$ & $0.798 \pm 0.078$ \\
& & 0  & $0.970 \pm 0.004$ & $0.987 \pm 0.001$ & $0.729 \pm 0.225$ & $0.009 \pm 0.000$ \\
& & 9  & $0.975 \pm 0.003$ & $0.985 \pm 0.000$ & $0.717 \pm 0.123$ & $0.100 \pm 0.002$ \\
& & 13 & $0.967 \pm 0.010$ & $0.985 \pm 0.001$ & $0.792 \pm 0.132$ & $0.083 \pm 0.011$ \\
\cmidrule{2-7}
& \multirow{4}{*}{1.0}
& --- & $0.975 \pm 0.003$ & $0.981 \pm 0.002$ & $0.862 \pm 0.062$ & $0.878 \pm 0.014$ \\
& & 0  & $0.977 \pm 0.002$ & $0.987 \pm 0.001$ & $0.945 \pm 0.044$ & $0.011 \pm 0.002$ \\
& & 9  & $0.974 \pm 0.005$ & $0.984 \pm 0.001$ & $0.968 \pm 0.020$ & $0.101 \pm 0.002$ \\
& & 13 & $0.976 \pm 0.005$ & $0.985 \pm 0.001$ & $0.964 \pm 0.026$ & $0.096 \pm 0.003$ \\

\midrule

\multirow{8}{*}{$\star$6}
& \multirow{4}{*}{$\star$0.5}
& --- & $0.967 \pm 0.007$ & $0.975 \pm 0.009$ & $0.502 \pm 0.274$ & $0.847 \pm 0.041$ \\
& & 0  & $0.973 \pm 0.005$ & $0.986 \pm 0.002$ & $0.924 \pm 0.046$ & $0.010 \pm 0.000$ \\
& & 9  & $0.967 \pm 0.009$ & $0.985 \pm 0.001$ & $0.956 \pm 0.037$ & $0.101 \pm 0.001$ \\
& & 13 & $0.974 \pm 0.006$ & $0.984 \pm 0.001$ & $0.890 \pm 0.045$ & $0.090 \pm 0.007$ \\
\cmidrule{2-7}
& \multirow{4}{*}{1.0}
& --- & $0.969 \pm 0.003$ & $0.980 \pm 0.004$ & $0.858 \pm 0.025$ & $0.881 \pm 0.034$ \\
& & 0  & $0.978 \pm 0.003$ & $0.983 \pm 0.006$ & $0.977 \pm 0.017$ & $0.029 \pm 0.008$ \\
& & 9  & $0.977 \pm 0.006$ & $0.985 \pm 0.001$ & $0.952 \pm 0.027$ & $0.099 \pm 0.001$ \\
& & 13 & $0.978 \pm 0.003$ & $0.984 \pm 0.001$ & $0.909 \pm 0.040$ & $0.092 \pm 0.004$ \\

\midrule

\multirow{8}{*}{8}
& \multirow{4}{*}{0.5}
& --- & $0.969 \pm 0.009$ & $0.982 \pm 0.002$ & $0.561 \pm 0.144$ & $0.761 \pm 0.124$ \\
& & 0  & $0.972 \pm 0.pa5$ & $0.989 \pm 0.000$ & $0.936 \pm 0.025$ & $0.010 \pm 0.000$ \\
& & 9  & $0.972 \pm 0.005$ & $0.985 \pm 0.001$ & $0.901 \pm 0.073$ & $0.100 \pm 0.002$ \\
& & 13 & $0.972 \pm 0.004$ & $0.984 \pm 0.002$ & $0.917 \pm 0.045$ & $0.091 \pm 0.005$ \\
\cmidrule{2-7}
& \multirow{4}{*}{1.0}
& --- & $0.972 \pm 0.007$ & $0.981 \pm 0.003$ & $0.893 \pm 0.029$ & $0.870 \pm 0.036$ \\
& & 0  & $0.975 \pm 0.004$ & $0.987 \pm 0.001$ & $0.953 \pm 0.037$ & $0.036 \pm 0.012$ \\
& & 9  & $0.977 \pm 0.004$ & $0.986 \pm 0.001$ & $0.956 \pm 0.035$ & $0.102 \pm 0.000$ \\
& & 13 & $0.977 \pm 0.002$ & $0.985 \pm 0.003$ & $0.957 \pm 0.018$ & $0.095 \pm 0.003$ \\

\bottomrule
\end{tabular}
}
\caption{WaNet ablation study (grid size and warp strength).}
\label{tab:wanet}
\end{table}

Specific ablation studies were conducted for each attack to find the most effective parameters. Nevertheless, for space reason we report only the analysis conducted to find the poisoning ratio (\Cref{tab:ablation}) and the WaNet parameters (\Cref{tab:wanet}). 
A validation set forged by using the $20\%$ of the training set has been used for this purpose. The selected attack parameters are indicated by $\star$.

\subsection{Clean Performances} 
\label{appendix:clean} 

\Cref{tab:clean_performance} reports the clean accuracies and F1-scores of the NN and DPL models over the four learning tasks.

\begin{table}[h] 
\centering \resizebox{0.95\columnwidth}{!}{%
\begin{tabular}{lcccc}
\toprule
\multirow{2}{*}{\textbf{Task}} &
\multicolumn{2}{c}{\textbf{NN}} &
\multicolumn{2}{c}{\textbf{DPL}} \\
\cmidrule(lr){2-3} \cmidrule(lr){4-5}
& \textbf{Accuracy} & \textbf{F1-macro} & \textbf{Accuracy} & \textbf{F1-macro} \\
\midrule
Addition       & $0.984 \pm 0.001$ & $0.983 \pm 0.001$ & $0.988 \pm 0.001$ & $0.988 \pm 0.001$ \\
Multiplication & $0.985 \pm 0.002$ & $0.983 \pm 0.003$ & $0.988 \pm 0.000$ & $0.987 \pm 0.001$ \\
Binding        & $0.927 \pm 0.005$ & $0.927 \pm 0.005$ & $0.932 \pm 0.001$ & $0.932 \pm 0.000$ \\
SDDOIA         & $0.852 \pm 0.002$ & $0.825 \pm 0.002$ & $0.848 \pm 0.005$ & $0.819 \pm 0.005$ \\
\bottomrule
\end{tabular}
}
\caption{Clean Performances.}
\label{tab:clean_performance}
\end{table}

\subsection{Autonomous Driving Logic Rules}
\label{appendix:rules} 

\noindent
\begin{minipage}{\columnwidth}
\centering

\noindent
\resizebox{\columnwidth}{!}{%
\begin{minipage}{\columnwidth} 
\begin{align*} 
\texttt{invalid\_fs}(I) &\leftarrow \texttt{go\_fs}(I) \land \texttt{stop\_fs}(I), \\
\texttt{invalid\_fs}(I) &\leftarrow \texttt{go\_fs}(I) \land \texttt{obstacle}(I), \\[0.3em] \texttt{move\_forward}(I) &\leftarrow \texttt{go\_fs}(I) \land \neg\texttt{invalid\_fs}(I) \land \neg\texttt{stop\_fs}(I), \\
\texttt{no\_stop}(I) &\leftarrow \texttt{go\_fs}(I) \land \neg\texttt{invalid\_fs}(I) \land \neg\texttt{stop\_fs}(I), \\
\texttt{no\_stop}(I) &\leftarrow \neg\texttt{go\_fs}(I) \land \neg\texttt{stop\_fs}(I), \\
\texttt{not\_move}(I) &\leftarrow \texttt{go\_fs}(I) \land \neg\texttt{invalid\_fs}(I) \land \texttt{stop\_fs}(I), \\
\texttt{not\_move}(I) &\leftarrow \neg\texttt{go\_fs}(I), \\
\texttt{stop}(I) &\leftarrow \texttt{go\_fs}(I) \land \neg\texttt{invalid\_fs}(I) \land \texttt{stop\_fs}(I), \\ 
\texttt{stop}(I) &\leftarrow \neg\texttt{go\_fs}(I) \land \texttt{stop\_fs}(I), \\[0.5em] 
\texttt{left}(I) &\leftarrow \texttt{go\_left}(I), \\
\texttt{left}(I) &\leftarrow \neg\texttt{any\_left}(I) \land \texttt{coin}(I,21), \\
\texttt{no\_left}(I) &\leftarrow \neg\texttt{go\_left}(I) \land \texttt{any\_left}(I), \\ 
\texttt{no\_left}(I) &\leftarrow \neg\texttt{any\_left}(I) \land \neg\texttt{coin}(I,21), \\[0.5em] 
\texttt{right}(I) &\leftarrow \texttt{go\_right}(I) \land \neg\texttt{block\_right}(I), \\ 
\texttt{right}(I) &\leftarrow \neg\texttt{any\_right}(I) \land \texttt{coin}(I,21), \\ 
\texttt{no\_right}(I) &\leftarrow \texttt{go\_right}(I) \land \texttt{block\_right}(I), \\ 
\texttt{no\_right}(I) &\leftarrow \neg\texttt{go\_right}(I) \land \texttt{any\_right}(I), \\ 
\texttt{no\_right}(I) &\leftarrow \neg\texttt{any\_right}(I) \land \neg\texttt{coin}(I,21), \\[0.5em]
\texttt{action}(I,a) &\leftarrow a(I), \qquad a \in \left\{ \begin{array}{c} \texttt{not\_move},\, \texttt{move\_forward},\, \texttt{no\_stop},\, \texttt{stop},\\ 
\texttt{no\_left},\, \texttt{left},\, \texttt{no\_right},\, \texttt{right} \end{array} \right\}.
\end{align*} 
\end{minipage} 
} 
\logicrule{Autonomous Driving. The \textsc{move\_forward} and \textsc{stop} actions share the same FS concept block, leading the target label \textsc{Forward + Stop} to be unsatisfiable.}
\label{sddoia_logic}
\end{minipage}

\subsection{FTrojan on MNIST Tasks}
\label{appendix:digit_classification}
The performances of the FTrojan attack on MNIST digit classification, Addition and Multiplication are reported in ~\Cref{tab:digit_classification}.
\begin{table}[h] 
\centering 
\resizebox{0.85\columnwidth}{!}{%
\begin{tabular}{lccc} 
\hline 
\textbf{Task} & \textbf{Accuracy} & \textbf{F1 Macro} & \textbf{\ac{ASR}} \\ 
\hline Classification & $0.983 \pm 0.004$ & $0.983 \pm 0.004$ & $0.604 \pm 0.403$ \\
Addition & $0.977 \pm 0.006$ & $0.981 \pm 0.004$ & $0.017 \pm 0.006$ \\
Multiplication & $0.979 \pm 0.005$ & $0.979 \pm 0.005$ & $0.013 \pm 0.006$ \\ 
\hline 
\end{tabular}%
}
\caption{Performance of FTrojan on the MNIST tasks.}
\label{tab:digit_classification}
\end{table}

\section{Ethics and Privacy Statement}
\begin{minipage}{0.5\textwidth}
    This work uses only publicly available benchmark datasets and does not process personal, sensitive, or user-generated data. Our experiments are conducted in controlled research settings and aim to characterize, rather than facilitate, attacks against neuro-symbolic systems. We disclose attack configurations to support reproducibility and the development of effective countermeasures. Nevertheless, the demonstrated techniques could be misused to compromise learning systems, and should therefore be applied only to authorized models and datasets.
\end{minipage}

\end{document}